%% file: main.tex
\documentclass[conference]{IEEEtran}
\IEEEoverridecommandlockouts
\usepackage{color}
\usepackage{tabularx}
\usepackage{epsfig}
\usepackage{graphicx}
\usepackage[T1]{fontenc}
\usepackage{epstopdf}
\usepackage{comment}
\makeatletter
\usepackage{xcolor}
\newcommand{\thickhline}{%
	\noalign {\ifnum 0=`}\fi \hrule height 1.3pt
	\futurelet \reserved@a \@xhline
}
\newcolumntype{"}{@{\hskip\tabcolsep\vrule width 1pt\hskip\tabcolsep}}
\makeatother
\usepackage{hyperref}
\usepackage{arydshln}
\usepackage[labelfont=bf, font=small]{caption}
\newcommand{\algorithmautorefname}{Algorithm}
\def\sectionautorefname~#1\null{Section #1\null}
\def\equationautorefname~#1\null{Equation (#1)\null}
\def\algorithmautorefname~#1\null{Algorithm #1\null}
\def\subsectionautorefname~#1\null{Section #1\null}
\def\subsubsectionautorefname~#1\null{Section #1\null}
\definecolor[named]{ACMDarkBlue}{cmyk}{1,0.58,0,0.21}
\definecolor{myblue}{rgb}{0.0, 0.2, 0.8}
\definecolor{mygray}{rgb}{0.7, 0.7, 0.7}
\hypersetup{
	colorlinks,
	linkcolor={myblue!100!black},
	citecolor={myblue!100!black},
	urlcolor={myblue!100!black}
}

\usepackage{algorithm} 
\usepackage{algpseudocode} 

\usepackage{textcomp}
\usepackage{xcolor}
\usepackage{cite}
\usepackage{amsmath,amssymb,amsfonts}
\def\BibTeX{{\rm B\kern-.05em{\sc i\kern-.025em b}\kern-.08em
    T\kern-.1667em\lower.7ex\hbox{E}\kern-.125emX}}

\begin{document}

\title{L2Fuzz: Discovering Bluetooth L2CAP Vulnerabilities Using Stateful Fuzz Testing}

\author{\IEEEauthorblockN{Haram Park, Carlos Kayembe Nkuba, Seunghoon Woo, Heejo Lee\IEEEauthorrefmark{1}}
\textit{Korea University}, \{freehr94, carlosnkuba, seunghoonwoo, heejo\}@korea.ac.kr}

\maketitle
\thispagestyle{plain}
\pagestyle{plain}

\input{0-abstract}
\input{1-intro}
\input{2-back_motiv}
\input{3-metho}

\input{4-eval}

\input{5-dis}
\input{6-rel-works}
\input{7-concl}

\section*{Acknowledgment}
We appreciate the anonymous reviewers and our shepherd for their
helpful comments. % to improve our paper. 
This work was
supported by Institute of Information \& Communications Technology Planning \& Evaluation (IITP) grant funded by the Korea
government (MSIT) (No.2019-0-01697 Development of Automated
Vulnerability Discovery Technologies for Blockchain Platform Security, 
No.2019-0-01343 Regional Strategic Industry Convergence
Security Core Talent Training Business, and No.IITP-2022-2020-0-01819 ICT Creative Consilience program).

\bibliographystyle{IEEEtran}
\bibliography{references}

% \newpage
% \onecolumn
\appendices

\input{8-appendix}

% \newpage

\end{document}

%% file: 0-abstract.tex
\begin{abstract}
Bluetooth Basic Rate/Enhanced Data Rate (BR/EDR) is a wireless technology used in billions of devices. Recently, several Bluetooth fuzzing studies have been conducted to detect vulnerabilities in Bluetooth devices, but they fall short of effectively generating malformed packets. In this paper, we propose \textsc{L2Fuzz}, a stateful fuzzer to detect vulnerabilities in Bluetooth BR/EDR Logical Link Control and Adaptation Protocol (L2CAP) layer. By selecting valid commands for each state and mutating only the core fields of packets, \textsc{L2Fuzz} can generate valid malformed packets that are less likely to be rejected by the target device. Our experimental results confirmed that: (1) \textsc{L2Fuzz} generates up to 46 times more malformed packets with a much less packet rejection ratio compared to the existing techniques, and (2) \textsc{L2Fuzz} detected five zero-day vulnerabilities from eight real-world Bluetooth devices. %, including a denial-of-service on Android where a CVE ID will be assigned. 
\end{abstract}
% (maximum 150-word). % two CVE (Android denial-of-service, Apple Dos) will be assigned.
\begin{IEEEkeywords}
Bluetooth, Fuzz Testing, Wireless Security.
\end{IEEEkeywords}

%% file: 1-intro.tex
\section{Introduction}
\label{sec:intro}

	\newcommand\blfootnote[1]{%
		\begingroup
		\renewcommand\thefootnote{}\footnote{#1}%
		\addtocounter{footnote}{-1}%
		\endgroup
	}
	
Bluetooth is a wireless communication technology that allows users to exchange various data in a short range, including Bluetooth Basic Rate/Enhanced Data Rate (BR/EDR) and Bluetooth Low Energy (BLE). Owing to their convenience, billions of devices have adopted Bluetooth technologies~\cite{bt_market}. Because Bluetooth is an open standard, most vendors install similar Bluetooth host stacks (\textit{i.e.}, software stack) in their devices for interoperability among different vendors. This also implies that it is easy for malicious users to craft wireless attacks on Bluetooth devices, and even a single vulnerability has the risk of being exploited in billions of devices.\blfootnote{* Heejo Lee is the corresponding author.}

To address such undesirable situations, several studies have been conducted to detect unknown Bluetooth vulnerabilities. However, existing studies are limited in their ability to perform fuzz testing in various Bluetooth devices; %In particular, 
they (1) required the Bluetooth pairing process (\textit{e.g.}, Defensics \cite{defensics}), (2) failed to generate valid malicious packets (\textit{e.g.}, BFuzz\cite{kim2017poster}), (3) did not consider state information (\textit{e.g.}, Bluetooth stack smasher \cite{bss}), or (4) were inefficient for testing various Bluetooth devices (\textit{e.g.}, KNOB \cite{antonioli2019knob}, BIAS \cite{antonioli2020bias}, BlueMirror \cite{claverie2021bluemirror}), all of which impair the effectiveness of Bluetooth fuzzing in terms of detecting critical vulnerabilities (see \autoref{sec:rel}).

To overcome these shortcomings, we propose \textsc{L2Fuzz}, a stateful fuzzer for Bluetooth host stacks, which targets the Logical Link Control and Adaptation Protocol (L2CAP) layer. Because all Bluetooth services use the L2CAP that is located in the lowest layer, L2CAP was chosen for our study in order to guarantee the root-of-trust of Bluetooth devices. 

\textsc{L2Fuzz} uses the following two key techniques: \textit{state guiding} and \textit{core field mutating}. Through \textit{state guiding}, \textsc{L2Fuzz} maps valid commands for each L2CAP state based on their events, functions and actions. Here, the mapped commands can be used for state transitions or to test valid attacks against a specific state. Next, \textsc{L2Fuzz} mutates only the core (\textit{i.e.}, critical) fields of L2CAP packets, which are in charge of the port and channel setting, through \textit{core field mutating} technique. By mutating only core fields while not modifying other parts, \textsc{L2Fuzz} can generate more valid test packets, resulting in more efficient detection of potential vulnerabilities. 

When we applied \textsc{L2Fuzz} to the selected eight test devices, \textsc{L2Fuzz} detected five zero-day vulnerabilities in three smartphones, one wireless earphone, and one laptop; we reported all detected vulnerabilities %, including denial-of-service attacks on Android devices (a CVE ID will be assigned), 
to the corresponding vendors.

To demonstrate the effectiveness of \textsc{L2Fuzz}, we compared it with existing Bluetooth fuzzing techniques~\cite{bss, kim2017poster, defensics}. To this end, we devised two novel metrics suitable for evaluating Bluetooth fuzzers: \textit{mutation efficiency} and \textit{state coverage}. The mutation efficiency includes the number of error-prone test packets that a fuzzer can generate, and the number of test packets that are rejected by the target. The state coverage represents the number of protocol states that a fuzzer can test.

From our experiments, we confirmed that \textsc{L2Fuzz} outperformed existing Bluetooth fuzzing techniques~\cite{bss, kim2017poster, defensics}. Compared with the existing techniques, L2Fuzz was able to: (1) cover and test up to 10 (out of 19) more L2CAP states, (2) generate up to 46 times more valid malformed packets, and (3) significantly reduce the probability that test packets would be rejected on the target device (up to 60\% reduction).

This paper makes the following three main contributions:

\begin{itemize}
\setlength\itemsep{0.3em}
\item %\textbf{\textit{Novel fuzzer.}} 
We present \textsc{L2Fuzz}, a stateful fuzzer for the Bluetooth host stack L2CAP layer. The key technical contributions is generating valid packets that are not likely to be rejected by the target device using \textit{state guiding} and \textit{core field mutating}. The source code of \textsc{L2Fuzz} is available at \url{https://github.com/haramel/L2Fuzz}.

\item %\textbf{\textit{New evaluation metrics.}} 
We devised two novel metrics that were suitable for evaluating Bluetooth fuzzers: \textit{mutation efficiency} and \textit{state coverage}, which can be used even in an environment where the target device is a black-box.

\item %\textbf{\textit{Validation of real-world devices.}}
When we applied \textsc{L2Fuzz} to eight Bluetooth devices, it detected five zero-day vulnerabilities, including a denial-of-service on Android devices %(a CVE will be assigned) 
and a crash on Apple devices.
\end{itemize}

%% file: 2-back_motiv.tex
\section{Background and Motivation}
\label{sec:bac}
In this section, we first introduce an overview of Bluetooth BR/EDR 5.2, mainly focusing on L2CAP, and then discuss the motivation behind the development of \textsc{L2Fuzz}. %\autoref{subsec:overview} provides an overview of Bluetooth BR/EDR 5.2, mainly focusing on L2CAP. \autoref{problem_statement} describes our target problems, and \autoref{motivation} motivates the \textsc{L2Fuzz} approach.

\subsection{Overview of Bluetooth BR/EDR and L2CAP}
\label{subsec:overview}

\begin{figure}[!ht]
\centering
    \includegraphics[width=\linewidth]{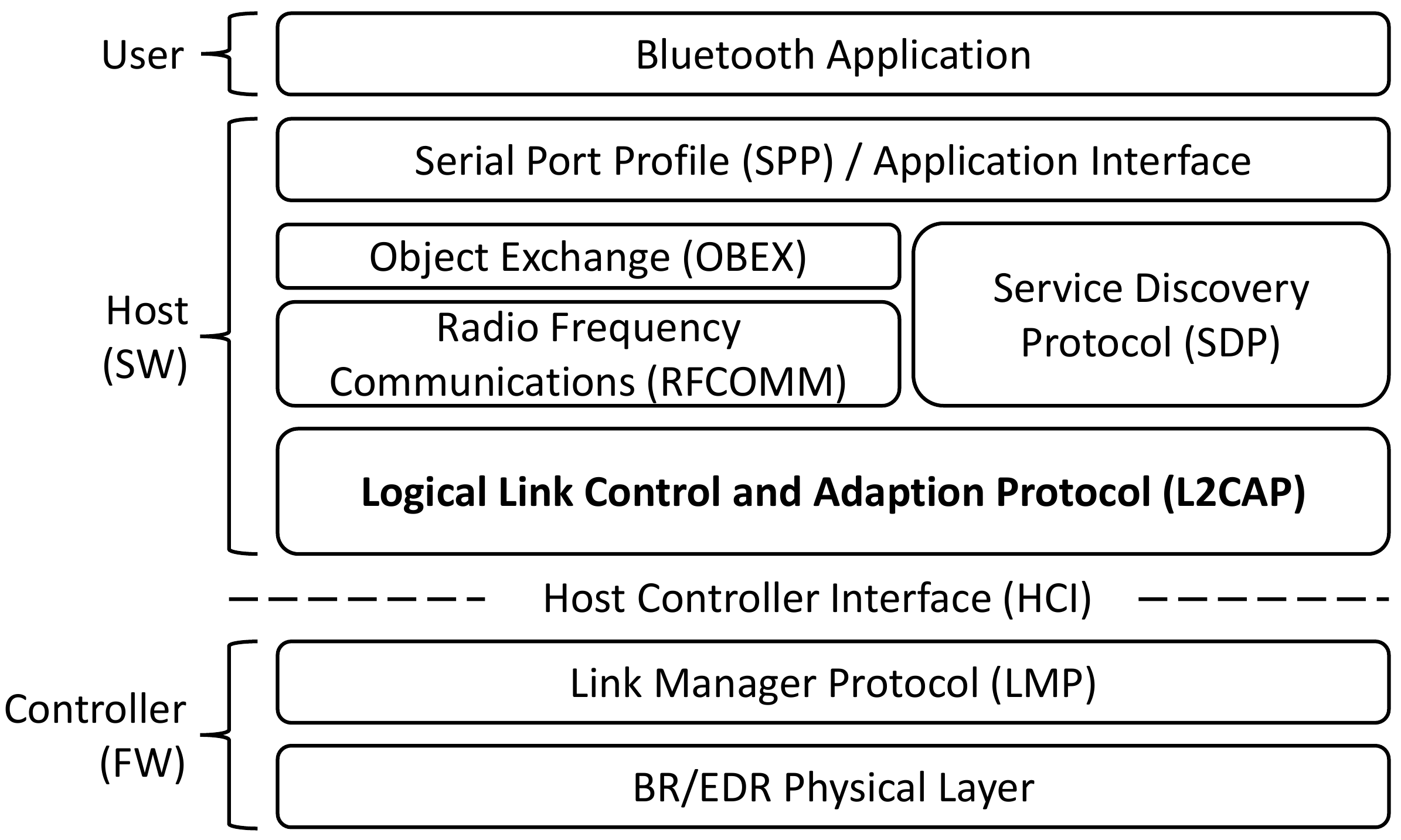}
\caption{Illustration for Bluetooth protocol stack.}
\label{fig:bt_stack}
\end{figure}

Bluetooth BR/EDR is a wireless technology for short-range communication \cite{bt_spec}. A Bluetooth network consists of a master device (\textit{i.e.}, initiator) and a slave device (\textit{i.e.}, acceptor). Both devices use a Bluetooth protocol stack,
which consists of a controller stack (\textit{i.e.}, firmware) and a host stack (\textit{i.e.}, software),
for communication (see \autoref{fig:bt_stack}). 
%for communication, as shown in \autoref{fig:bt_stack}. The Bluetooth protocol stack consists of a controller stack (\textit{i.e.}, firmware) and a host stack (\textit{i.e.}, software). 
Bluetooth BR/EDR is also referred to as Bluetooth in this paper. To use the service provided by the Bluetooth application, the master and slave devices should start the pairing process using their controller stack. Then, both devices create a connection between the L2CAP layer, which is the lowest layer of the host stack, enabling the connection between the upper layers. Since L2CAP is the lowest layer of the host stack, secure use of higher-layer protocols in Bluetooth applications requires a security assessment of L2CAP to ensure a root of trust.

%\vspace{6px}
\textbf{Introduction to L2CAP.} 
L2CAP is a core and essential protocol in Bluetooth because all Bluetooth applications require an L2CAP connection between the master and slave devices~\cite{hua2008analysis, sharan2018air}. To use Bluetooth applications, the master must know the service ports and channels of the slave services, which are handled by L2CAP. For example, %when we use a Bluetooth file transfer service,
suppose we intend to use a Bluetooth file transfer service. During this process, the master and slave devices first exchange an encrypted key using the controller stack. Thereafter, they share service ports and channels through the L2CAP layer. Based on these ports and channels, they create Radio Frequency Communications (RFCOMM) and Object Exchange (OBEX) connections to use file transfer applications.

\begin{figure}[t]
\centering
    \includegraphics[width=\linewidth]{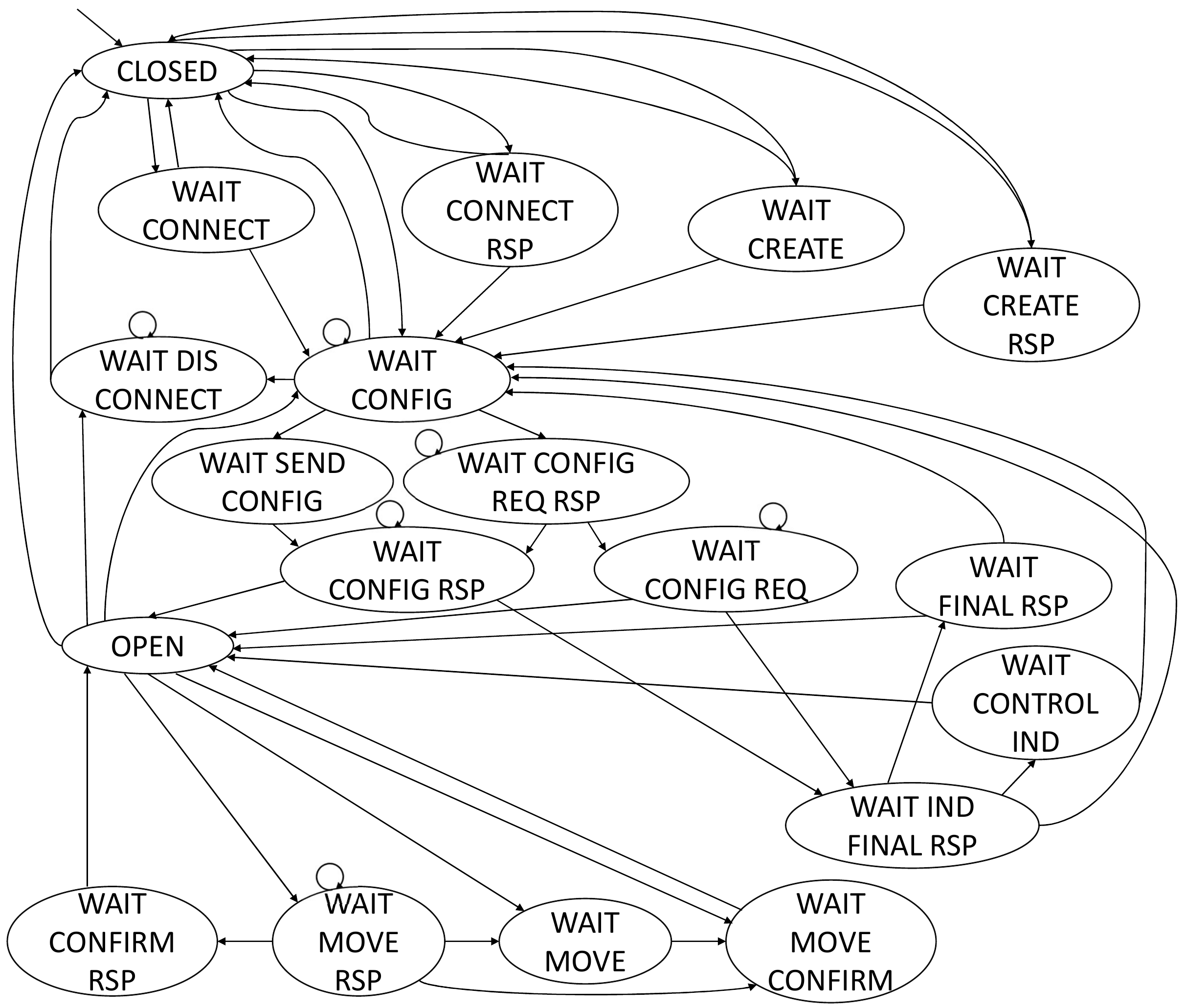}
\caption{Bluetooth 5.2 L2CAP state machine.}
\label{fig:l2cap_state_machine}
\end{figure}

\begin{figure}[t]
\centering
    \includegraphics[width=\linewidth]{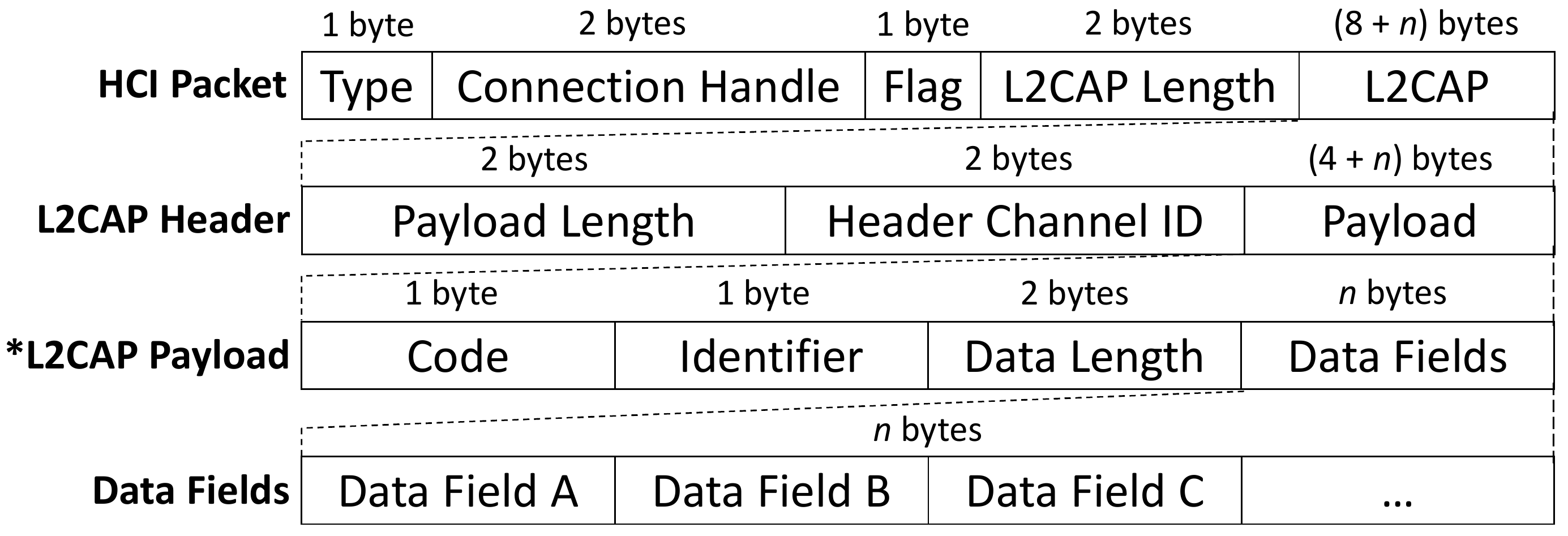}\\
    \footnotesize{*L2CAP Payload can be up to 65,535 bytes.}
\caption{Bluetooth L2CAP protocol format.}
\label{fig:l2cap_format}
\end{figure}

%\vspace{6px}
\textbf{L2CAP states.}
L2CAP is based on the concept of channels and consists of 19 states in Bluetooth 5.2 (see \autoref{fig:l2cap_state_machine}). These states cover various scenarios that can occur during the L2CAP communication process. Each state has \textit{event} and \textit{action} for the state transition. To transit from the current state to the next state, a specific request should be sent to the device (\textit{i.e.}, \textit{event}) and an appropriate response to that request should be output from the device (\textit{i.e.}, \textit{action}).

%\vspace{6px}
\textbf{L2CAP format.}
L2CAP consists of a header and a payload. The payload can have multiple \texttt{Data Fields} depending on the command being used (see \autoref{fig:l2cap_format}). To transmit an L2CAP packet, the packet must be configured in the order of an HCI packet, an L2CAP header, and an L2CAP payload. 

The HCI packet is used to communicate between the host and the controller. The L2CAP header consists of a \texttt{Payload Length} and \texttt{Header Channel ID}; the \texttt{Header Channel ID} identifies the destination channel endpoint of the packet and transmits L2CAP commands using the signaling channel (\textit{i.e.}, 0x0001). The L2CAP payload consists of \texttt{Code}, \texttt{Identifier}, \texttt{Data Length}, and \texttt{Data Fields}. \texttt{Code} and \texttt{Identifier} indicate the command code and packet ID, respectively. Next, \texttt{Data Fields} vary depending on the L2CAP command; there are 26 L2CAP commands in Bluetooth 5.2, and each command has different \texttt{Data Fields}. For example, ``L2CAP connection request'' has two \texttt{Data Fields}: Protocol/Service Multiplexer (PSM, port number) and Source Channel ID (SCID). Meanwhile, ``L2CAP connection response'' has four \texttt{Data Fields}: Destination Channel ID (i.e., DCID), SCID, Result and Status.

\subsection{Problem Statement} 
\label{problem_statement}
In this paper, we focus on detecting L2CAP vulnerabilities in a Bluetooth host stack. Because L2CAP is the lowest layer of the host stack, it is necessary to ensure a root of trust.
In particular, L2CAP vulnerabilities (\textit{e.g.}, BlueBorne~\cite{blueborne} and SweynTooth~\cite{garbelini2020sweyntooth}) can compromise the security of the entire system, such as causing denial-of-service and remote code execution. Therefore, to improve the security of Bluetooth applications, an effective technique that can detect vulnerabilities in the L2CAP layer is required.

Several existing approaches have attempted to resolve such undesirable situations; in particular, fuzz testing (i.e., fuzzing) was mainly performed for security verification of wireless communications~\cite{defensics, kim2017poster, bss}. However, they are limited in detecting potential vulnerabilities in Bluetooth applications owing to the following two main challenges. Failure to overcome these challenges can collectively impair the effectiveness of fuzzing.

%\vspace{6px}
\textbf{Challenge 1: Increasing the L2CAP state coverage.}
Because Bluetooth is a stateful protocol, it follows a specific state machine (see \autoref{fig:l2cap_state_machine}) and moves from one state to another~\cite{satam2017bluetooth}, each of which contains its own functions to perform the desired operation (\textit{e.g.}, a connection-related operation is conducted in the WAIT CONNECT state). The L2CAP provides state-transition functions to enter each state. 

Because vulnerabilities are highly likely to occur in (1) the state transition process and (2) the functions of each state, we need to validate the security of as many states as possible. However, increasing the state coverage of a fuzzer is difficult owing to the complexity of the protocol and various implementations of the Bluetooth stack. In fact, implementations using Bluetooth specifications are conducted differently according to the vendors preferences. Therefore, it is challenging to cover several L2CAP states on such a diverse implementation of Bluetooth applications (\textit{e.g.}, BSS~\cite{bss} can cover only three out of 19 L2CAP states).

%\vspace{6px}
\textbf{Challenge 2: Generating valid malformed packets.}
A malformed packet refers to a packet wrapped with malicious information or data~\cite{prabadevi2014distributed}. 
%Becuase
Because malformed packets have a higher chance of causing crashes that lead to fatal vulnerabilities (\textit{e.g.}, denial-of-service and buffer overflows)~\cite{dunning2010taming, patel2012survey, yang2015cybersecurity}, an effective Bluetooth fuzzer needs to generate valid malformed packets that are not rejected by the target device. 

However, the method of simply mutating any or all fields of L2CAP packets without considering the characteristics of each field, which is used in the existing Bluetooth fuzzing techniques, results in most of the generated packets being rejected by the target devices~\cite{torres2020nfdfuzz}.

% \textcolor{red}{In addition, it is difficult to apply the mutating algorithm of the existing file fuzzing studies (\textit{e.g.}, AFL \cite{afl}) to Bluetooth fuzzing. This is because there are various types of Bluetooth stacks as many as the number of manufacturers, and most of them are not open sources.}
In addition, it is challenging to apply the existing mutating algorithms of file fuzzing studies (\textit{e.g.}, AFL \cite{afl}) to Bluetooth fuzzing; because there are as many Bluetooth stacks as the number of manufacturers, and most of them are not open-sourced, the technology of existing file fuzzing studies cannot be applied.

\subsection{A Motivating Example}
\label{motivation}
\begin{figure} [t] %[h!]
\centering
    \includegraphics[width=\linewidth]{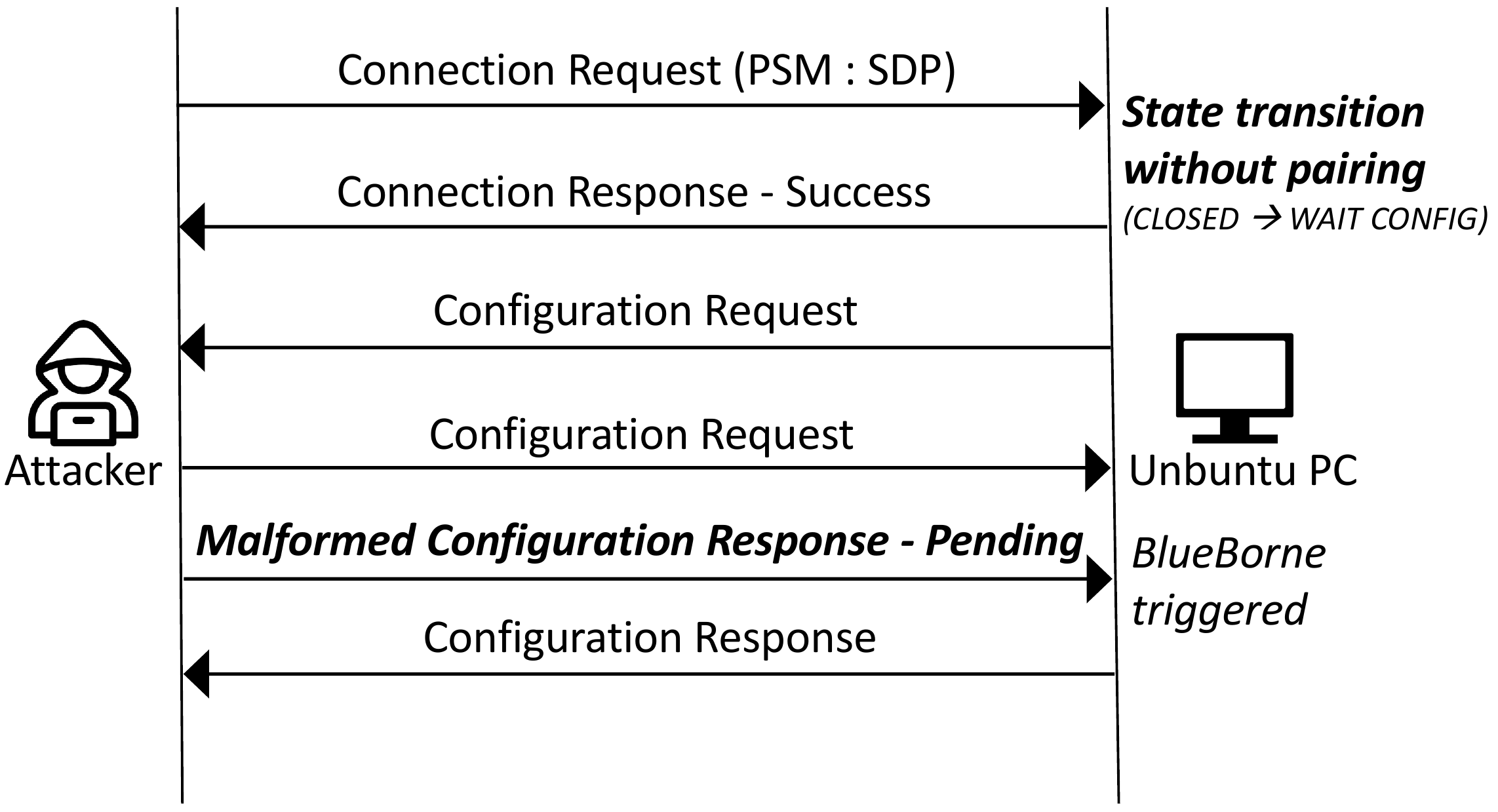}
\caption{Illustration of CVE-2017-1000251 attack flow.}
\label{fig:blueborne}
\end{figure}

To suggest the need for an effective Bluetooth vulnerability detection technique,
we introduce the CVE-2017-1000251 case (called  BlueBorne~\cite{blueborne}), which is a critical remote code execution vulnerability discovered in the Linux Bluetooth host stack (\textit{i.e.}, BlueZ~\cite{bluez}). \autoref{fig:blueborne} illustrates the attack flow of the BlueBorne.

First, the attacker forms an L2CAP connection using a service discovery protocol (SDP) port that does not require pairing. After the L2CAP connection is established, the victim device enters the configuration state. During the configuration process, the attacker sends a normal configuration request packet and a malformed configuration response packet to the victim device. Because these packets are valid in the configuration state, the victim's device accepts the packets without rejection, which leads to a fatal attack.

In the overall attack flow, we focused on the two main steps that are central to the BlueBorne attack scenario: entering the configuration state, and sending malformed packets. Existing Bluetooth fuzzing techniques: (1) do not fully consider the L2CAP states, and (2) do not generate valid malformed packets (for testing purposes) for each L2CAP state. Therefore, they easily fail to reach the specific state (e.g., configuration state), and even if they reach it, they send invalid packets that are rejected by the target device, which results in a failure to identify critical vulnerabilities including BlueBorne.

In this regard, a mature fuzzing technique is required to overcome the two aforementioned challenges (see \autoref{problem_statement}), that is, to generate malicious packets valid for each state while having high L2CAP state coverage.

%% file: 3-metho.tex
\section{Methodology}
\label{sec:met}
In this section, we introduce the methodology of \textsc{L2Fuzz}, a stateful fuzzer for detecting Bluetooth L2CAP vulnerabilities.

\subsection{Overview}
\label{sub:l2fuzz_overview}
\begin{figure}[t] %[h!]
\centering
\includegraphics[width=\linewidth]{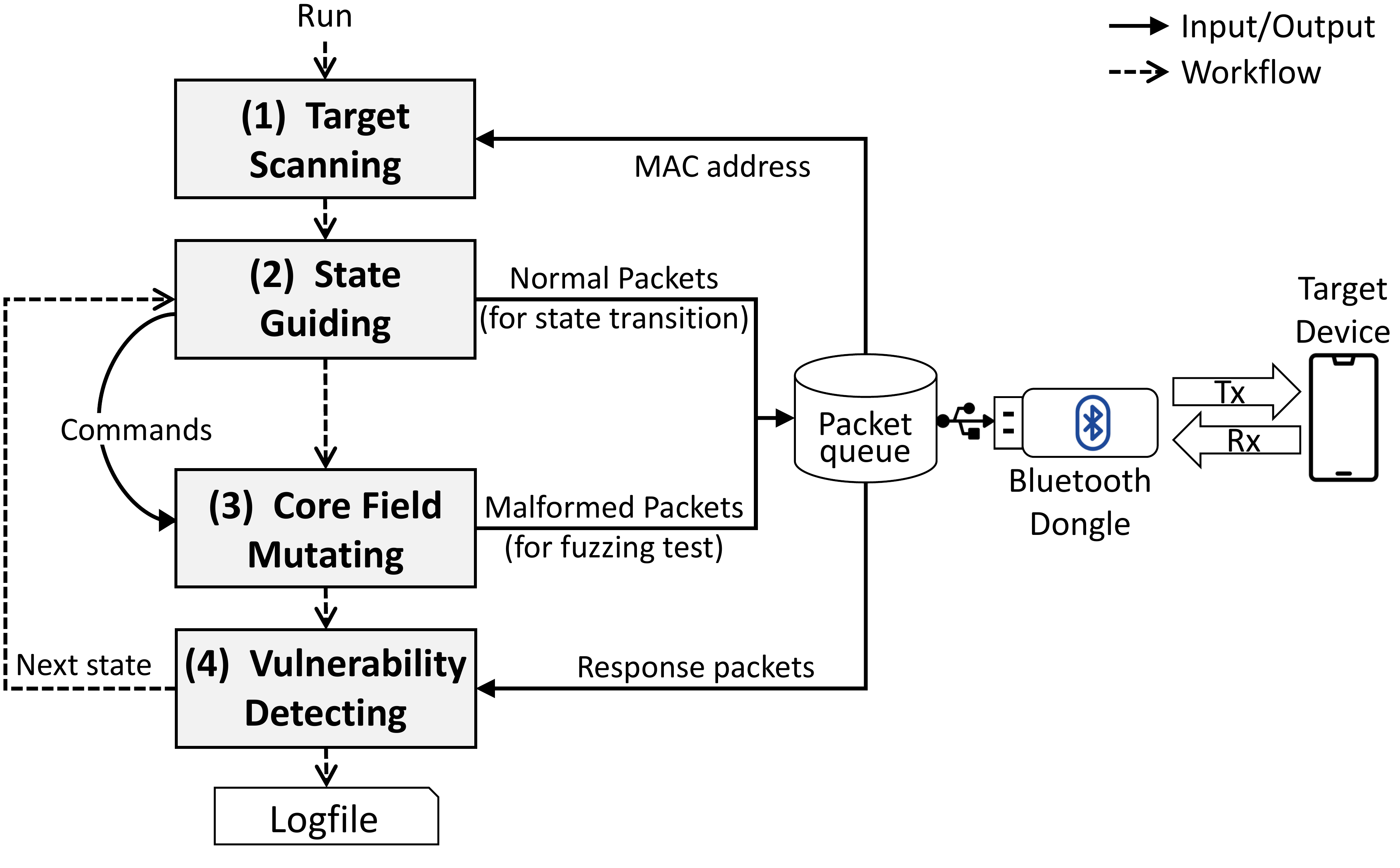}
\vspace{-1.5em}    
\caption{High-level workflow of \textsc{L2Fuzz}.}
\label{fig:l2fuzz_overview}
\end{figure}

\autoref{fig:l2fuzz_overview} depicts the overall workflow of \textsc{L2Fuzz}, which comprises the following four phases: (1) target scanning, (2) state guiding, (3) core field mutating, and (4) vulnerability detecting. \textsc{L2Fuzz} first scans a target Bluetooth device and creates an L2CAP socket with its MAC address by connecting a port that does not require pairing. \textsc{L2Fuzz} then traverses L2CAP states and generates malformed packets for testing purposes. For state transition, \textsc{L2Fuzz} uses predefined valid commands for each of the 19 L2CAP states; these valid commands were mapped based on the event and action of each state. To generate valid malformed packets, \textsc{L2Fuzz} segments an L2CAP packet format into parts to be mutated (\textit{i.e.}, mutable fields) and parts to be maintained. By mutating only mutable fields while it does not change the maintained parts, \textsc{L2Fuzz} can generate valid malformed packets that are less likely to be rejected in the target device. Finally, \textsc{L2Fuzz} detects potential vulnerabilities by sending the generated malformed packets to the target device.

\subsection{Target scanning}\label{phase1}
\textsc{L2Fuzz} first scans the target device's meta-information, namely, the MAC address (for establishing L2CAP socket), device name, class of the device (\textit{e.g.}, smartphone or laptop), and its Organizationally Unique Identifier (OUI). Thereafter, \textsc{L2Fuzz} scans the service ports of the target device to detect ports that do not require pairing. Because (1) attackers often exploit Bluetooth vulnerabilities without pairing, (\textit{e.g.}, see \autoref{motivation}) (2) fuzzing after pairing is meaningless as this process is the same as remotely controlling the target device after gaining permission, and (3) for ports that require pairing, sending test packets without pairing causes the device to reject command packets without parsing any fields. Therefore, we decided that L2Fuzz should perform fuzz testing without pairing to counteract such external attacks.

%\vspace{6px}
\textbf{Potentially exploitable port scanning.} 
To find potentially exploitable ports that do not require pairing, \textsc{L2Fuzz} receives a list of supported service ports on the target device and attempts to connect to each service port. If all ports are identified as requiring pairing, \textsc{L2Fuzz} then attempts to connect to the SDP port, which does not require pairing and is supported by every Bluetooth device~\cite{shafranovich2015bluetooth, jungbluemaster} as an alternative. 
By communicating with the target device through the port that does not require pairing, 
\textsc{L2Fuzz} can detect L2CAP vulnerabilities that are not detected in a paired situation.
% 중복표현이라 제거 
% Various L2CAP vulnerabilities, including the BlueBorne vulnerability, attempt attacks without a pairing process by leveraging a port that does not require pairing. L2Fuzz can detect L2CAP vulnerabilities that are not detected in a paired situation.

\subsection{State guiding}
\label{phase2}
Next, \textsc{L2Fuzz} traverses each L2CAP state in the target device. To increase the state coverage, we first identify valid commands for each state, and use them in state transition.

% Next, \textsc{L2Fuzz} identifies valid commands for L2CAP states to successfully proceed with the state transition and generates a valid test packet later.

%\vspace{6px}
\textbf{State classification.}
Before identifying valid commands, we first clustered the 19 L2CAP states based on the \textit{event} that each state receives, the \textit{functions} that internally process the desired operation in each state, and the corresponding output \textit{action}, which we referred to as a \textit{job}. Consequently, we classified the 19 L2CAP states into the following seven jobs:

\begin{table} [h]
\centering
\renewcommand{\tabcolsep}{2.4mm}
\caption{Jobs categorized based on events, functions, and actions of L2CAP states.}
\label{table:wait_con}
\scalebox{1}{
\begin{tabular}{cl}
\thickhline
\rule{0in}{2.2ex}\textbf{Job} & \textbf{States}\\
\hline
\rule{0in}{2.2ex}Closed & \{CLOSED\}\\\hdashline[2pt/2pt]
\rule{0in}{2.5ex}Connection & \{WAIT CONNECT, WAIT CONNECT RSP\}\\\hdashline[2pt/2pt]
\rule{0in}{2.5ex}Creation & \{WAIT CREATE, WAIT CREATE RSP\}\\\hdashline[2pt/2pt]
\rule{0in}{2.5ex}Configuration & \begin{tabular}[l]{@{}l@{}}\rule{0in}{2.5ex}\{WAIT CONFIG, WAIT CONFIG RSP,\\$\;\,$WAIT CONFIG REQ, WAIT CONFIG REQ RSP,\\$\;\,$WAIT SEND CONFIG, WAIT IND FINAL RSP,\\$\;\,$WAIT FINAL RSP, WAIT CONTROL IND\}\end{tabular}\\\hdashline[2pt/2pt]
\rule{0in}{2.5ex}Disconnection & \{WAIT DISCONNECT\}\\\hdashline[2pt/2pt]
\rule{0in}{2.5ex}Move & 
\begin{tabular}[l]{@{}l@{}}\rule{0in}{2.5ex}\{WAIT MOVE, WAIT MOVE RSP,\\$\;\,$WAIT MOVE CONFIRM, WAIT CONFIRM RSP\}\end{tabular}\\\hdashline[2pt/2pt]
\rule{0in}{2.5ex}Open & \{OPEN\}\\
%\hline
\thickhline
\end{tabular}
}
\end{table}

By identifying the commands used for each job, we included commands in the test packet, which are valid for the state of the target device; this significantly reduces the possibility that the packet will be rejected by the target device. Moreover, because valid commands can be shared between different states in a job, more diverse test packets can be generated; consequently, fuzzing coverage can be increased.

As an example of job classification, we introduce the process of ``Connection job'' identification. \autoref{table:wait_con} summarizes the events and actions of the WAIT CONNECT state confirmed in the Bluetooth 5.2 specification document.

\begin{table} [!h]
\centering
\renewcommand{\tabcolsep}{3mm}
\caption{WAIT CONNECT state's events and actions.}
\label{table:wait_con}
\scalebox{1}{
\begin{tabular}{llc}
\thickhline
\rule{0in}{2.5ex}\textbf{Event} & \textbf{Action} & \textbf{State transition?}\\
\hline
\rule{0in}{2.5ex}Connect Req & Connect Rsp & WAIT CONFIG\\\hdashline[2pt/2pt]
\rule{0in}{2.5ex}Connect Rsp & Reject & No \\
Config Req & Reject & No \\
Config Rsp & Reject & No \\
Disconnect Rsp & Reject & No \\
Create Channel Req & Reject & No \\
Create Channel Rsp & Reject & No \\
Move Channel Req & Reject & No \\
Move Channel Rsp & Reject & No \\
Move Channel Confirm Req & Reject & No \\
Move Channel Confirm Rsp & Reject & No \\
%\hline
\thickhline
\end{tabular}
}
\end{table}

When the target device is in the WAIT CONNECT state, if we send a connection request (Connect Req) packet, the target device does not reject the packet because this packet is valid in the current state. After executing the function related to the connection, the target device sends a corresponding response packet (Connect Rsp) and %transitions
changes its state to WAIT CONFIG. We confirmed that in the WAIT CONNECT RSP state, the target device performed almost similar operations (\textit{i.e.}, events, functions, and actions); thus, we classified WAIT CONNECT and WAIT CONNECT RSP as the \textit{connection job}.

One consideration was that Bluetooth devices did not always display the exact same operations as defined in the documentation. For example, some Android devices did not reject the ``Connect Rsp'' event even though the device was in the WAIT CONNECT state. This occurred because of the variety of implementations of the Bluetooth stack. Therefore, we set the boundaries of valid commands for each job slightly more generously to increase fuzzing effectiveness, knowing that several packets may be rejected.

Consequently, we map valid commands to each job based on the specification and various packet traces. The valid commands mapped for each job are shown in \autoref{table:mapping}.

\begin{table} [h]
\centering
\renewcommand{\tabcolsep}{4mm}
\caption{Valid commands mapped for each job.}
\label{table:mapping}
\scalebox{1}{
\begin{tabular}{cl}
\thickhline
\rule{0in}{2.2ex}\textbf{Job} & \textbf{Valid commands}\\
\hline
\rule{0in}{2.2ex}Closed & All commands\\\hdashline[2pt/2pt]
\rule{0in}{2.5ex}Connection & Connect Req/Rsp\\\hdashline[2pt/2pt]
\rule{0in}{2.5ex}Creation & Create Channel Req/Rsp\\\hdashline[2pt/2pt]
\rule{0in}{2.5ex}Configuration & Config Req/Rsp\\\hdashline[2pt/2pt]
\rule{0in}{2.5ex}Disconnection & Disconnect Req/Rsp\\\hdashline[2pt/2pt]
\rule{0in}{2.5ex}Move &
\begin{tabular}[l]{@{}l@{}}\rule{0in}{2.5ex}Move Channel Req/Rsp,\\
Move Channel Confirmation Req/Rsp\end{tabular}\\\hdashline[2pt/2pt]
\rule{0in}{2.5ex}Open & All commands\\
\thickhline
\end{tabular}
}
\end{table}

By clustering states into jobs and mapping valid commands for each job, \textsc{L2Fuzz} can cover most of the L2CAP states in the security validation of Bluetooth devices while decreasing the test packet rejection rate. This, in turn, renders \textsc{L2Fuzz} more likely to detect potential L2CAP vulnerabilities.

%\vspace{6px}
\textbf{State transition.}
With the valid commands, \textsc{L2Fuzz} generates normal packets, and then sends the packets to the target device through the packet queue. After receiving the packets, the target device enters the corresponding state and sends a response packet. The packet queue parses the response packets and returns the state transition result. If \textsc{L2Fuzz} succeeds in the state transition, it obtains the valid commands for the target state and generates a valid malformed packet in next phase. 
When the fuzz testing of the target state is completed, the state transition is executed again to move to the next target state.

\subsection{Core field mutating}
\label{phase3}
\textsc{L2Fuzz} then generates malformed packets that can lead to vulnerabilities in the entered L2CAP state of the target device. To increase the effectiveness of fuzzing, the generated packets should not be rejected by the target device.

To this end,
we decided not to mutate fields that can easily be checked for anomalies. In particular, we segmented an L2CAP packet format into parts to be mutated (\textit{i.e.}, mutable fields) and parts to be maintained by referring to \textsc{Vfuzz}~\cite{nkuba2021riding}. 

%\vspace{6px}
\textbf{Field classification.}
Let $L$ be the L2CAP packet. We segment $L$ into fixed ($F$), dependent ($D$), and mutable ($M$) fields as follows:
\begin{equation*}\label{eq1}
L = F \cup D \cup M    
\end{equation*}

\begin{itemize}
  \item $F$ is a set of \textit{fixed fields}. The values are fixed.
  \item $D$ is a set of \textit{dependent fields}.
  This values are determined by other values.
  \item $M$ is a set of \textit{mutable fields}. The values are determined by devices or users.
\end{itemize}

%\vspace{6px}
We further classified $M$ into mutable \textit{core} fields ($M_C$) and mutable \textit{application} fields ($M_A$) to distinguish only core fields that can affect the core functions of the L2CAP:
\begin{equation*}\label{eq2}
M = M_C \cup M_A
\end{equation*}

\begin{itemize}
    \setlength\itemsep{0.3em}
    \item $M_C$ is a set of \textit{mutable core fields}. The values determine the port and channel for Bluetooth network.
    \item $M_A$ is a set of \textit{mutable application fields}. The values vary depending on commands and convey data to the target.
\end{itemize}

\begin{figure}[t]
\centering
    \includegraphics[width=\linewidth]{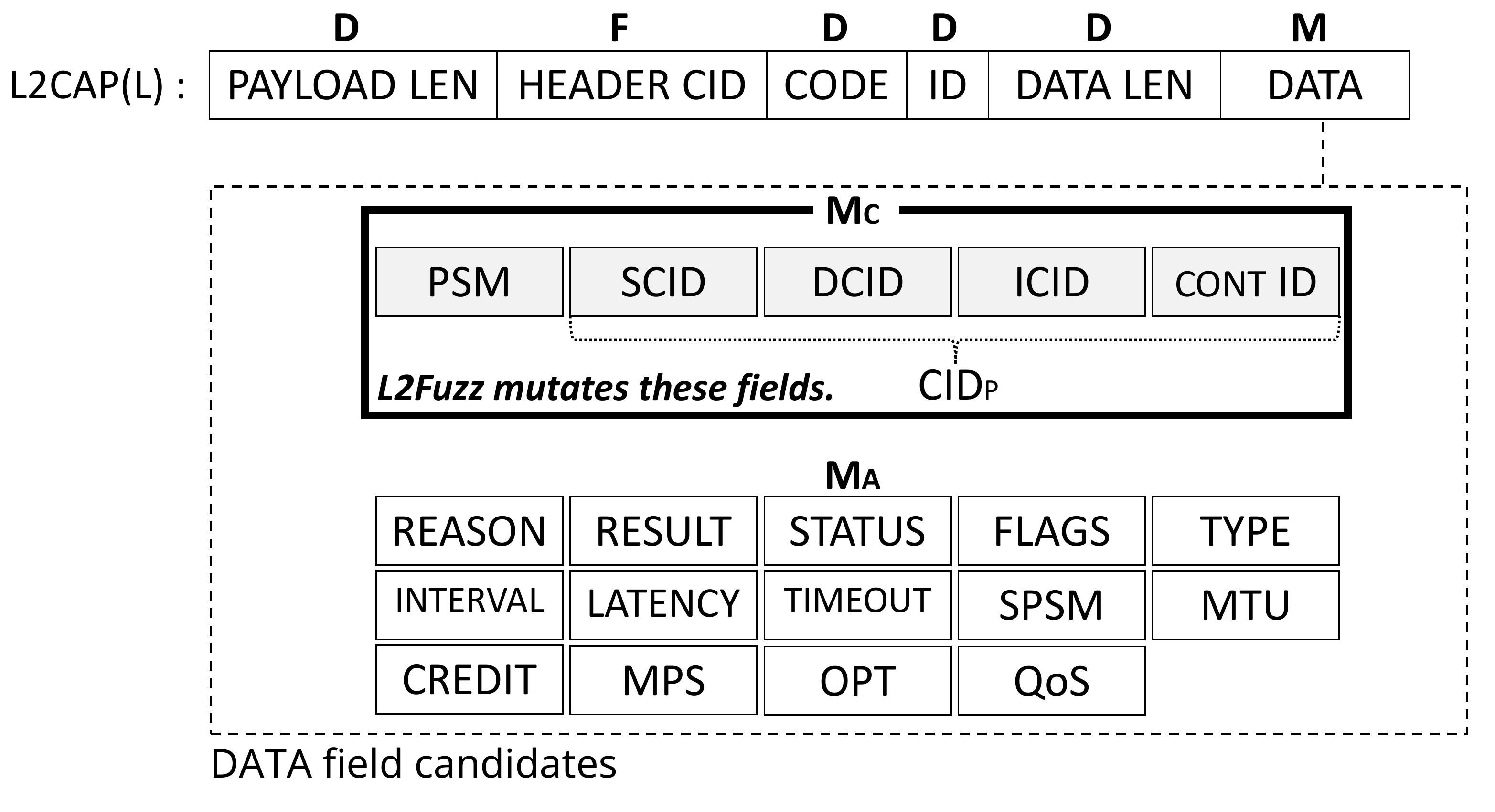}
\caption{Field classification for Bluetooth 5.2 L2CAP packet frame.}
\label{fig:l2cap_frame}
\vspace{-1em}
\end{figure}

%\vspace{6px}

Consequently, an L2CAP packet $L$ can be conceptually segmented as follows:
$L = F \cup D \cup  M_C \cup M_A$.
Based on this concept, we classified each field of the Bluetooth 5.2 L2CAP frame structure (see \autoref{fig:l2cap_frame}) as follows.
%The detailed descriptions are as follows.
 
%\vspace{6px}
\begin{itemize}
    \setlength\itemsep{0.8em}
    \item \textbf{$\boldsymbol{F}$ = \{HEADER CID\}}. 
    \begin{itemize}
        \item[*] This field is fixed because \texttt{0x0001} is used to manage the channel over the ACL-U logical links.
    \end{itemize}
    \item $\boldsymbol{D}$ = \textbf{\{PAYLOAD LEN, CODE, ID, DATA LEN\}}. 
    \begin{itemize}
        \item[*] PAYLOAD LEN is determined by the length of the information payload (CODE, ID, DATA LEN, and DATA), CODE is determined by the valid command code (\textit{i.e.}, determined in the state guiding phase), ID is dynamically assigned by the device, and DATA LEN is determined by the length of DATA.
    \end{itemize}
    \item $\boldsymbol{M_C}$ = \textbf{\{PSM, SCID, DCID, ICID, CONT ID\}}. 
      \begin{itemize}
        \item[*] PSM is used for port settings. SCID, DCID, ICID, and CONT ID are responsible for setting the channel endpoint, and % SCID, DCID, ICID, and CONT ID 
        are also referred to as ``Channel ID in Payload ($\mathrm{CID_P}$)'' in this paper.
      \end{itemize}
    \item $\boldsymbol{M_A}$ = \textbf{\{REASON, RESULT, STATUS, FLAGS, TYPE, INTERVAL, LATENCY, TIMEOUT, SPSM, MTU, CREDIT, MPS, OPT, QoS\}}.
    \begin{itemize}
        \item[*] Each value contains application data for commands, and is intended to deliver data without affecting port or channel management.
      \end{itemize}        
 \end{itemize}

%\vspace{6px}
\textbf{Packet mutation.} 
To minimize packet rejection by the target device, \textsc{L2Fuzz} mutates only the $M_C$ field. 

\textsc{L2Fuzz} does not mutate $F$ and $D$ to avoid possible packet rejection. If the target device receives a packet with $F$ or $D$ mutated, it will send a reject command response for ``Command not understood''. Furthermore, \textsc{L2Fuzz} maintains $M_A$ with its default values. This field is optional, thus, does not have a significant effect on the target device. Additionally, some of the fields can have up to 65,535 bytes of data, requiring a large amount of time to test various cases. Thus, \textsc{L2Fuzz} leaves these fields as default values.

\begin{table}[t] %[!h]
\centering
\caption{Range of $M_C$ that can be used as malicious data.}
\label{table:mt_table}
\renewcommand{\tabcolsep}{2.8mm}
\begin{tabular}{clll}
\thickhline
\rule{0in}{2.2ex}\textbf{Fields ($\boldsymbol{M_C}$)} &  &\textbf{Range (Hex)} &\\
\hline\rule{0in}{2.2ex}
& 0100 - 01FF & 0300 - 03FF & 0500 - 05FF \\
PSM & 0700 - 07FF & 0900 - 09FF & 0B00 - 0BFF \\
& 0D00 - 0DFF & All even values \\
\hdashline[2pt/2pt]\rule{0in}{2.2ex}
$\mathrm{CID_P}$ & 0040 - FFFF & & \\
\thickhline
\end{tabular}
\vspace{-0.5em}
\end{table}

In contrast, \textsc{L2Fuzz} mutates $M_C$ to generate various malformed packets. Specifically, different approaches are used for PSM and $\mathrm{CID_P}$ in the $M_C$ field. Regarding PSM (\textit{i.e.}, port number), its normal range is defined in the Bluetooth specification document, and each device supports service ports within this range. Notably, the normal range has already been tested when scanning ports in the target scanning phase. Therefore, \textsc{L2Fuzz} considers values belonging to their abnormal range (see \autoref{table:mt_table}) and proceeds with a packet mutation. Next, in the case of $\mathrm{CID_P}$ (\textit{i.e.}, SCID, DCID, ICID, and CONT ID), the value is dynamically assigned by the device during normal communication within the available range. If the target device receives an abnormal $\mathrm{CID_P}$ value, it sends a command rejection response for the reason ``Invalid CID in request'' Therefore, we decided to consider the normal range of $\mathrm{CID_P}$ (see \autoref{table:mt_table}) while ignoring dynamic allocation because, although the value is contained in the normal range, it can cause unexpected behavior on the target device due to ignoring dynamic allocation and putting different values. 

Finally, \textsc{L2Fuzz} appends a garbage value to the tail of the packet, which increases the possibility of vulnerability detection when the packet is not rejected but parsed by the target device. Here, we considered garbage values that do not exceed the MTU size; if the garbage value exceeds the MTU size, the target device rejects the packet with the reason ``Signaling MTU exceeded.''

\begin{figure}[t]
\centering
    \includegraphics[width=\linewidth]{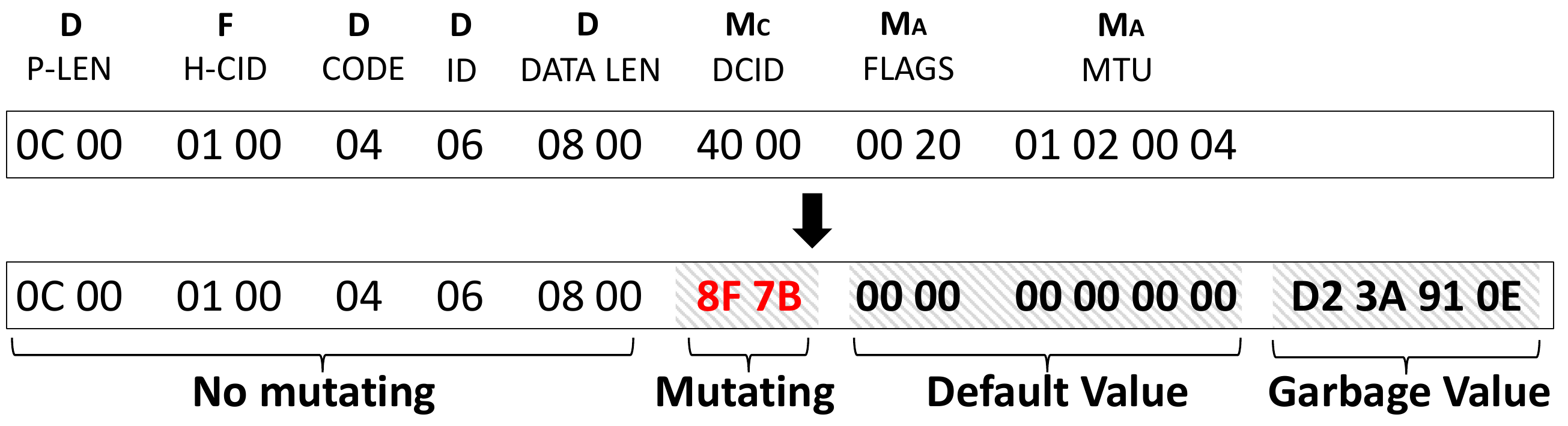}
\caption{Example of mutating L2CAP Config Req packet.}
\label{fig:mutating_example}

\vspace{1em}
\end{figure}

\autoref{fig:mutating_example} shows an example of mutating an L2CAP Config Req packet. \textsc{L2Fuzz} forcibly mutates the dynamically allocated DCID value (\textit{i.e.}, ``40 00'') into ``8F 7B'', and adds a garbage value (\textit{i.e.}, ``D2 3A 91 0E'') to the tail of the packet to generate a malformed packet.

\begin{algorithm}[t]
	\caption{Algorithm for core field mutating} 
	\label{alg:mutation}
	\hspace*{\algorithmicindent} \textbf{Input: $C$} \Comment{Valid commands (\textit{e.g.}, Connect Req)}
    \hspace*{\algorithmicindent} \textbf{Output: $pkt$\\}
    \hspace*{\algorithmicindent} \textcolor{mygray}{// Malformed packets to be stored in \textsc{packetQueue}}
    %\hspace{-5em}\Comment{Malformed packets, storing in \textsc{packetQueue}}
	\begin{algorithmic}[1]
        \Procedure{\textcolor{myblue}{\textbf{CoreFieldMutating}}}{$C$}
        \State{\textcolor{mygray}{// $n$: the number of malformed packets to generate}}
        \For{$i$ \textbf{in} range (0, \texttt{len}($C$))}
            \Comment{$i^{th}$ command}
            \For{$j$ \textbf{in} range (0, $n$)}
                \State{$pkt$ $\leftarrow$ \texttt{format}($commands[i]$)}
                \State{\textcolor{mygray}{// Fields: $F$, $D$, $M_C$, and $M_A$}}
                \State{$pkt.F$ $\leftarrow$ $0\text{x}0001$}
                \State{$pkt.D$ $\leftarrow$ \texttt{default}} \Comment{Used without changes}
                %\Comment{ACL-U}
                % \State{$pkt.D.value$}
                % \Comment{Depend on other values}
                \If{($pkt.M_C$)}
                    \If{$pkt.M_C.PSM$}
                        \State{\hspace{-1em}$pkt.M_C.PSM$ $\leftarrow$ \texttt{random}($abnormal$)}
                    \EndIf
                    \If{$pkt.M_C.CID_P$}
                        \State{\hspace{-1em}$pkt.M_T.CID_P$ $\leftarrow$ \texttt{random}($normal$)}
                    \EndIf
                \EndIf
                \If{$pkt.M_A$}
                    \State{$pkt.M_A$ $\leftarrow$ \texttt{default}}
                    \State{\textcolor{mygray}{// Used without changes}}
                \EndIf
                \State{$pkt$ = $pkt$.\texttt{append}($garbage$)}
                \State{\textsc{packetQueue}.\texttt{append}($pkt$)}
            \EndFor
        \EndFor
	   \EndProcedure
	\end{algorithmic} 
\end{algorithm}

Malformed packets generated in this manner are less likely to be rejected by the target device; thus, Bluetooth vulnerabilities can be detected more effectively than dumb mutation, which simply changes any or all fields of an L2CAP packet.
The generated malformed packets are transmitted to the target device through the packet queue. 
The high-level algorithm of this phase is explained in \autoref{alg:mutation}.

\begin{table*}[htbp] %[h!]
\centering
\caption{Summary of test devices used in the experiments.}
\vspace{-0.5em}
\label{table:devices}
\scalebox{0.97}{
\begin{tabular}{llllllllll}
  \thickhline
  \rule{0in}{2.2ex}
  \textbf{No.}& \textbf{Type} & \textbf{Vendor} & \textbf{Name} & \textbf{Year} &  \textbf{Model} & \textbf{Chip} & \textbf{OS or FW} & \textbf{BT Stack} & \textbf{BT Ver.} \\
  \hline\rule{0in}{2.2ex}%
  D1 & Tablet PC & Google & Nexus 7 & 2013  & ASUS-1A005A & Snapdragon 600 & Android 6.0.1  & BlueDroid & 4.0 + LE\\
  D2 & Smartphone & Google & Pixel 3 & 2018  & GA00464 & Snapdragon 845 & Android 11.0.1 & BlueDroid & 5.0 + LE \\
  D3 & Smartphone & Samsung & Galaxy 7 & 2016  & SM-G930L & Exynos 8890 & Android 8.0.0 & BlueDroid & 4.2 \\
  D4 & Smartphone & Apple & iPhone 6S & 2015  & A1688 & A9 & iOS 15.0.2 & iOS stack & 4.2 \\
  D5 & Earphone & Apple & Airpods 1 gen & 2016  & A1523 & W1 & 6.8.8 & RTKit stack & 4.2 \\
  D6 & Earphone & Samsung & Galaxy Buds+ & 2020  & SM-R175NZKATUR & BCM43015 & R175XXU0AUG1  & BTW & 5.0 + LE\\
  D7 & Laptop & LG & Gram & 2019  & 15ZD990-VX50K & Intel wireless BT & Windows 10 & Windows stack & 5.0 \\
  D8 & Laptop & LG & Gram & 2017  & 15ZD970-GX55K & Intel wireless BT & Ubuntu 18.04.4 & BlueZ & 5.0 \\
  \thickhline
\end{tabular}
}
\vspace{-1.2em}
\end{table*}

\subsection{Vulnerability detecting}
\label{phase4}
Finally, \textsc{L2Fuzz} detects the L2CAP vulnerabilities in the target device. To this end, \textsc{L2Fuzz} checks (1) whether the packet received at the target device contains an error message, (2) whether the ping test was successfully performed, and (3) whether a crash dump was generated.

Because L2CAP is concerned with port and channel settings, if a vulnerability is found, we receive an error message related to the Bluetooth connection, which can be one of the following: \textit{Connection Failed, Connection Aborted, Connection Reset, Connection Refused, and Timeout}. Of these, the \textit{Connection Failed} error means that the target Bluetooth service has been shut down, which can lead to a DoS. The remaining errors indicate a target device crash and may induce a crash in the target device.

Thereafter, \textsc{L2Fuzz} conducts a ping test. If the ping test fails, it is logged as a vulnerability according to the error message. \textsc{L2Fuzz} further checks whether there are any crash dumps or abnormalities. Finally, \textsc{L2Fuzz} stores the fuzzing results in a log file. 
When this step is over, the vulnerability assessment for the target state entered in the second phase (\textit{i.e.}, state guiding) is finished. Hence, after this step, \textsc{L2Fuzz} returns to the second phase, and assess the next L2CAP state.

After evaluating the vulnerabilities for all L2CAP states, the vulnerability detection process for the target device is finished, and \textsc{L2Fuzz} reports all the detected vulnerabilities.

%% file: 4-eval.tex
%\vspace{0.5em}
\section{Evaluation}
\label{sec:eval}
In this section, we evaluate \textsc{L2Fuzz}. \autoref{subsec:sut} introduces the experimental setup, including the experimental environment, target devices, and evaluation metrics. \autoref{vul_disco} investigates the vulnerability detection results of \textsc{L2Fuzz} on real-world Bluetooth devices. We then compare \textsc{L2Fuzz} with existing Bluetooth fuzzing techniques using the two metrics, mutation efficiency (\autoref{mut_eff}) and state coverage (\autoref{state_cov}), to demonstrate the effectiveness of \textsc{L2Fuzz}. Finally, \autoref{case_study} introduces a case study of the denial of service vulnerability detected in Android Bluetooth stack.
%\smallskip

\subsection{Experimental setup}\label{subsec:sut}

\textbf{Experiment environment.} 
\textsc{L2Fuzz} was implemented in approximately 1,200 lines of Python code, excluding external libraries. In particular, we used the Scapy library (v2.4.4), an interactive packet manipulation program for mutating packets. We ran \textsc{L2Fuzz} on a virtual machine with Ubuntu 18.04.4 LTS, 8GB memory, Intel Core i5-7500 CPU @ 3.40GHz × 4, 50GB Disk and Billionton Bluetooth Class 1 dongle. 

%\vspace{6px}
\textbf{Target devices.} 
We selected eight real-world test devices that can represent the general-purpose Bluetooth protocol stacks~\cite{bluetooth_stack}, including BlueZ (Linux), BlueDroid (Android), Apple BT stack, and Windows BT stack. \autoref{table:devices} summarizes the target devices used in the experiments.

%\noindent
%(Synopsys)
%(IoTcube)
%(SecuObs)
%\vspace{6px}
\textbf{Baseline fuzzers for comparison.}
When evaluating the effectiveness of \textsc{L2Fuzz}, we compared the results of \textsc{L2Fuzz} with those of Defensics~\cite{defensics}, BFuzz~\cite{kim2017poster}, and BSS~\cite{bss}; other related techniques were excluded because they did not support L2CAP vulnerability detection or were not publicly available. We compared the mutation efficiency and state coverage of \textsc{L2Fuzz} to the baseline fuzzers using the test device D2 (\textit{i.e.}, Google Pixel 3 smartphone, see \autoref{table:devices}). We used D2 in the evaluation because D2 follows the Bluetooth standard with little customization as a "reference phone" selected by Google. Therefore, we expected that Bluetooth vulnerabilities would be most clearly tested in D2.

%\noindent
%\vspace{6px}
\textbf{Evaluation metrics.} 
Because most Bluetooth stacks, except for BlueZ and BlueDroid, are closed sources, Bluetooth fuzzers are close to blackbox fuzzers; evaluation metrics used in whitebox or greybox fuzzing~\cite{afl, muench2018you}, such as source code coverage, are difficult to use for evaluation here. 
%Thus, we suggest two metrics for evaluating Bluetooth fuzzers:
%These two metrics can be measured only with the packet trace.
Thus, we suggest two metrics, which can be measured only with the packet trace, for evaluating Bluetooth fuzzers: \textit{mutation efficiency} and \textit{state coverage}, which can be measured even in an environment where the target device is a black-box:

\vspace{0.4em}
\begin{itemize}

%\item \textbf{Metric 1: Mutation efficiency.} 
%by the target device.
\item \textbf{Mutation efficiency.} 
This refers to the minimum percentage of malformed packets transmitted without rejection. To measure this metric, we calculated the \textit{Malformed Packet Ratio (MP Ratio)} and the \textit{Packet Rejection Ratio (PR Ratio)}, by capturing malformed and rejected packets through packet sniffing tools (\textit{e.g.}, Wireshark~\cite{lamping2004wireshark}).

\vspace{-0.5em}
\begin{equation*} \label{eq3}
\small
\begin{split}
\textit{MP Ratio} = \frac{\textit{\#Transmitted Malformed Packets}}{\textit{\#Transmitted Packets}}
\end{split}
\end{equation*}

\vspace{-0.5em}
\begin{equation*} \label{eq4}
\small
\begin{split}
\textit{PR Ratio} = \frac{\textit{\#Received Rejection Packets from Target}}{\textit{\#Received Packets from Target}}
\end{split}
\end{equation*}

\vspace{3px}
The mutation efficiency, which represents the ratio of malformed packets transmitted without rejection, is calculated as follows.

\vspace{-0.9em}
\begin{equation*} \label{eq5}
\small
\begin{split}
\textit{Mutation efficiency} = \textit{MP Ratio} * (1 - \textit{PR Ratio})
\end{split}
\end{equation*}

\item \textbf{State coverage.}
This metric refers to the number of L2CAP states to be covered. Because vulnerabilities are highly likely to occur in the state transition process and the functions of each state, the more L2CAP states were covered, the higher the likelihood of detecting vulnerabilities. It can be measured by protocol reverse engineering tool (\textit{e.g.}, PRETT \cite{lee2018prett}).
  
\end{itemize} 

\subsection{Vulnerability detection results in real-world devices}
\label{vul_disco}
We applied \textsc{L2Fuzz} to eight selected test devices for detecting unknown Bluetooth vulnerabilities. Owing to the characteristics of Bluetooth, wherein the device and fuzzing are terminated when a valid vulnerability is found, it is difficult to measure the number of detected vulnerabilities. Therefore, we measured (1) whether vulnerabilities were detected and (2) the elapsed time required to detect vulnerabilities.
% Owing to the characteristics of Bluetooth, wherein the device and fuzzing are terminated when a valid vulnerability is found, it is difficult to measure the number of detected vulnerabilities. Hence, we evaluated whether vulnerability was detected and further investigated the elapsed time required to detect the vulnerability.

In our experiments, we confirmed that \textsc{L2Fuzz} detected \textbf{five zero-day vulnerabilities}; the results are shown in \autoref{table:sum_vd}.

\begin{table}[t]
\vspace{1em}
\begin{center}
\caption{Vulnerability detection results of \textsc{L2Fuzz}.}
\renewcommand{\tabcolsep}{1.4mm}
\label{table:sum_vd}
\scalebox{1}{
\begin{tabular}{ccccc}
\thickhline
\textbf{Device} & \textbf{Vuln?} & \textbf{Description} & \textbf{Elapsed Time} & \textbf{Reported to Vendors?}\\
\hline
D1 & \textbf{Yes} & DoS & 1 m 32 s & Yes\\
D2 & \textbf{Yes} & DoS & 1 m 25 s & Yes\\
D3 & \textbf{Yes} & DoS & 7 m 11 s & Yes\\
D4 & No & \textcolor{gray}{\textsc{n/a}} & \textcolor{gray}{\textsc{n/a}} & \textcolor{gray}{\textsc{n/a}}\\
D5 & \textbf{Yes} & Crash & 40 s & Yes\\
D6 & No & \textcolor{gray}{\textsc{n/a}} & \textcolor{gray}{\textsc{n/a}} & \textcolor{gray}{\textsc{n/a}}\\
D7 & No & \textcolor{gray}{\textsc{n/a}} & \textcolor{gray}{\textsc{n/a}} & \textcolor{gray}{\textsc{n/a}}\\
D8 & \textbf{Yes} & Crash & 2 h 40 m & Discussing\\
\thickhline
\smallskip
\end{tabular}}
\end{center}
\vspace{-1.5em}
\footnotesize{* D1, D2, D3: A denial of service was triggered because of a null pointer dereference by malformed packets. The vendor became aware of this vulnerability.
%They will be patched as a result of our reporting, and one CVE ID will be assigned 
(see \autoref{case_study}).}\\

\vspace{-0.5em}
\footnotesize{* D5: The device was unexpectedly terminated owing to the malformed packets. This has been patched by the vendor.}\\

\vspace{-0.5em}
\footnotesize{* D8: A crash dump was generated owing to a general protection failure by malformed packets. We are discussing this issue with the vendor.}

\vspace{1em}
\end{table}

\textsc{L2Fuzz} discovered DoS vulnerabilities in three Android devices (\textit{i.e.}, D1, D2, and D3). The crash was triggered in the state of device, which allowed malicious commands with the value $\mathrm{CID_P}$. Additionally, a tombstone file (\textit{i.e.}, Android crash dump~\cite{android_tombstone}) was generated in each device, resulting in Bluetooth termination for all devices (DoS triggered); details are explained in \autoref{case_study}.

\textsc{L2Fuzz} further detected crashes in two devices (\textit{i.e.}, D5 and D8), a wireless earphone, and a laptop. Regarding D5, a crash occurred in a state that allowed commands with a malicious PSM value, resulting in an abnormal phenomenon (\textit{i.e.}, termination without any control). For D8, a crash dump file was created within the target device, and the Bluetooth communication content and general protection errors were recorded in the crash dump.

%The source codes of all Bluetooth stacks were not disclosed; therefore, it was infeasible to analyze them closely.
Notably, with the exception of D8, all vulnerabilities were detected within several minutes.
It was infeasible to closely analyze the direct factors affecting performance, because the source code of all Bluetooth stacks were not publicly disclosed.
Instead, we confirmed that the vulnerability was detected within one minute in D5 (supporting six service ports) while requiring more than two hours on D8 (supporting 13 service ports). Subsequently, we can infer that the elapsed time was determined based on the number of service ports provided and the logic complexity of Bluetooth applications. We responsibly reported all five detected vulnerabilities to the corresponding vendors.

Although \textsc{L2Fuzz} discovered five zero-day vulnerabilities, it failed to detect vulnerabilities in three devices: D4, D6, and D7, which used iOS, BTW, and Windows stack, respectively. Their Bluetooth stack is based on the Bluetooth specification document; however, they also have proprietary protocol layers and logic. They may have implemented an exception handling logic for malformed packets generated by \textsc{L2Fuzz}.

\subsection{Mutation efficiency measurement} 
\label{mut_eff}
Next, we measured the mutation efficiency of \textsc{L2Fuzz} and compared it with the three existing Bluetooth fuzzing techniques (\textit{i.e.}, Defensics, BSS, and BFuzz). For a fair comparison, controlled experiments were required. Because each fuzzer sends a different number of packets per second, we measured the MP and PR Ratios of each fuzzer based on 100,000 sent packets. We ran the four fuzzers on Google Pixel 3 with Android 11 devices (\textit{i.e.}, D2). Malformed and rejected packets were captured and analyzed using Wireshark.

\begin{figure}[t]
\centering
    \includegraphics[width=\linewidth]{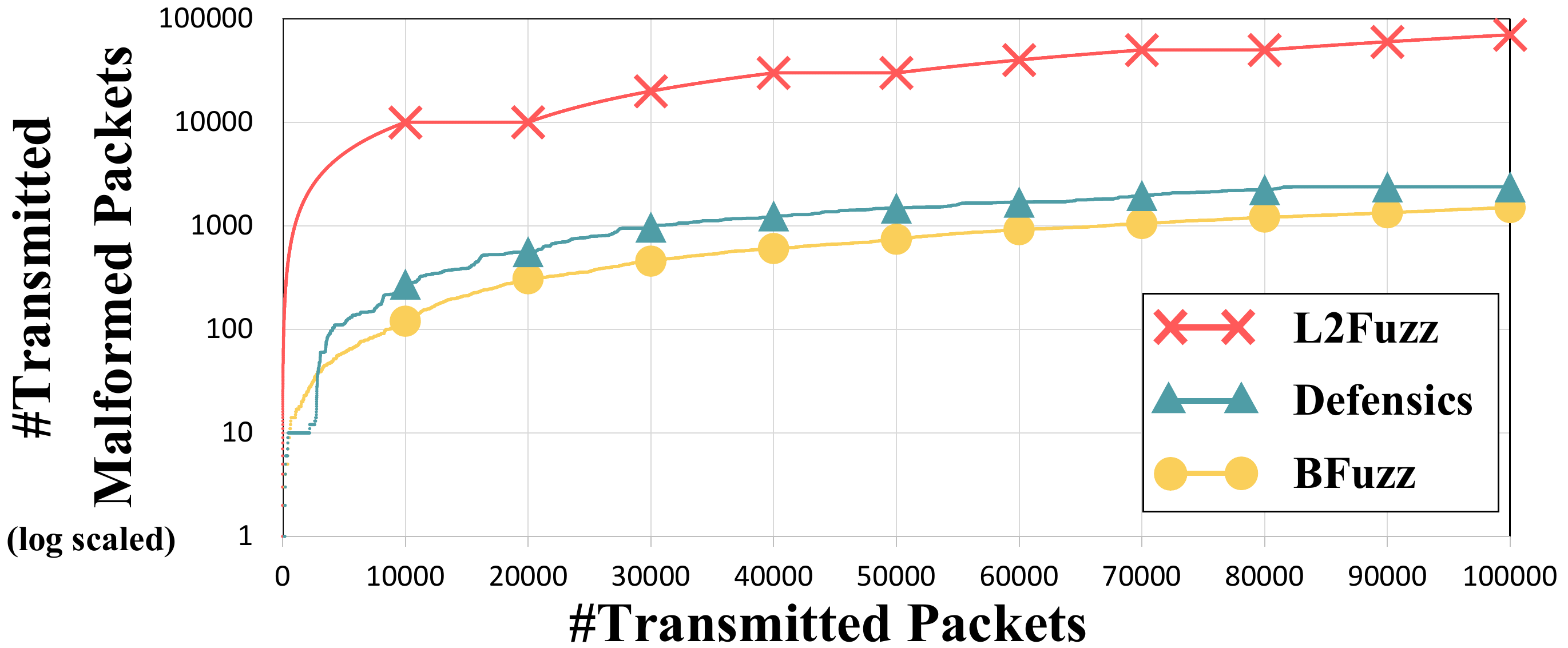}
\vspace{-1em}
\caption{MP Ratio measurement results for the four Bluetooth fuzzing techniques. BSS did not generate malformed packets,
thus it is not displayed on the graph.}
\label{fig:mal_graph}
\vspace{-1em}
\end{figure}

%\vspace{6px}
\textbf{MP Ratio measurement.}
\autoref{fig:mal_graph} shows the MP Ratio measurement results for the four fuzzing techniques. Notably, \textsc{L2Fuzz} can generate up to 46 times more malformed packets than other techniques. \textsc{L2Fuzz} generated malformed packets accounting for an average of 33.48\% during testing, and generated a total of 69,966 packets (\textit{i.e.}, 69.96\% \textit{MP Ratio}). Coversely, Defensics generated malformed packets accounting for 1.40\% on average and generated 2,380 packets in total (\textit{i.e.}, 2.38\% \textit{MP Ratio}). BFuzz generated malformed packets accounting for 0.74\% on average and generated a total of 1,506 packets (\textit{i.e.}, 1.50\% \textit{MP Ratio}). Notably, the BSS did not generate any malformed packets (\textit{i.e.}, 0\% \textit{MP Ratio}). 

We confirmed that the MP Ratio values of the fuzzing techniques vary depending on the mutation strategy used for each technique. Particularly, the existing techniques performed packet mutation without considering the characteristics of the L2CAP packet fields. For examples, BFuzz mutated all fields of the packet except for the fixed fields, and BSS mutated only one field. Therefore, they failed to effectively generate malformed packets. However, the \textsc{L2Fuzz} approach, which generates malformed packets with core field mutating, showed much higher MP Ratio than others.

\begin{figure}[t]
\centering
    \includegraphics[width=\linewidth]{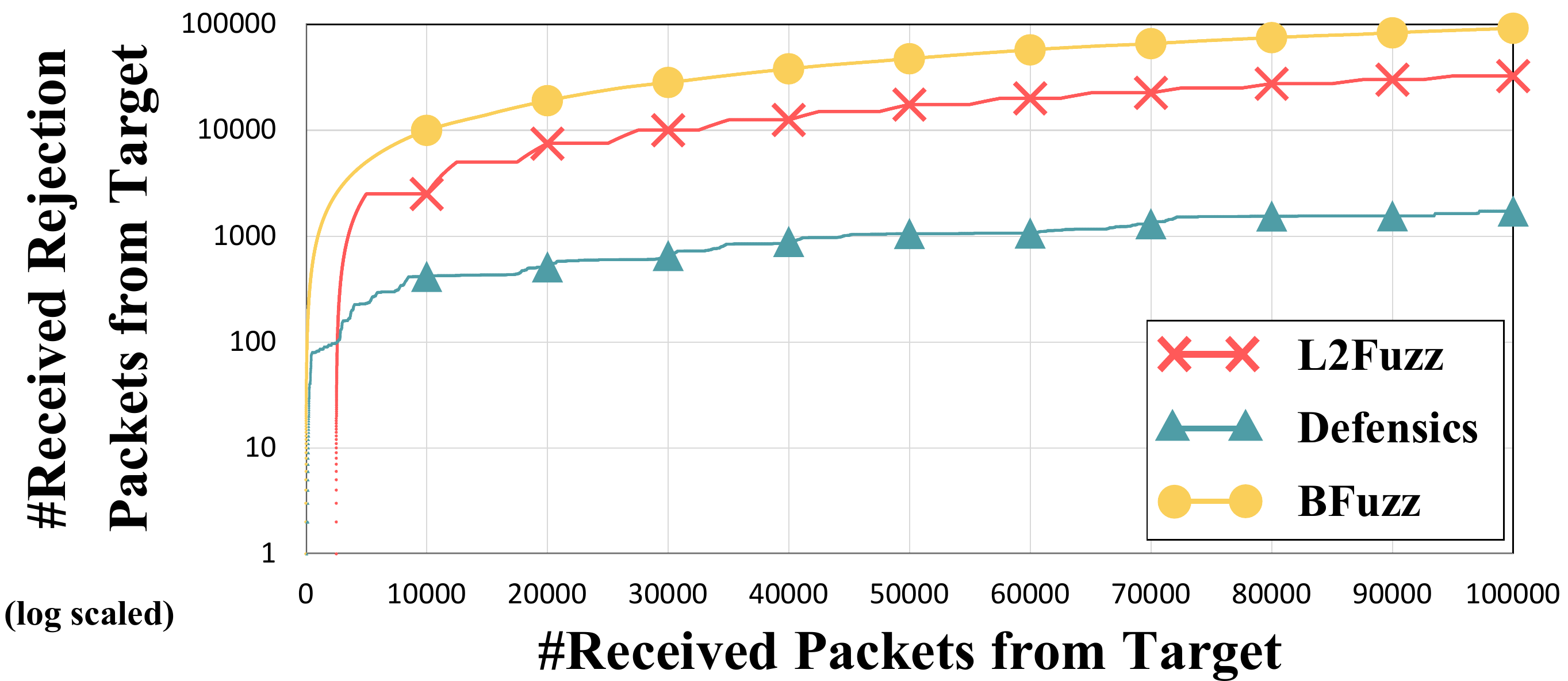}
\vspace{-0.3em}
\caption{PR Ratio measurement results for the four Bluetooth fuzzing techniques. BSS did not receive any rejection packets, thus it is not displayed on the graph.} \label{fig:rej_graph}

\vspace{0.5em}
\end{figure}

%\vspace{6px}
\textbf{PR Ratio measurement.}
\autoref{fig:rej_graph} shows the measured PR Ratio values for the selected four fuzzing techniques. Based on 100,000 received packets from the target, BFuzz showed the highest packet rejection ratio (\textit{i.e.}, 91.60\% \textit{PR Ratio}), followed by L2Fuzz (\textit{i.e.}, 32.49\% \textit{PR Ratio}), and then Defensics (\textit{i.e.}, 1.73\% \textit{PR Ratio}). Because BSS does not generate malicious packets, the PR Ratio was also 0\%.

Similar to the MP Ratio cases,
the PR Ratio values were different owing to the difference in the mutation strategy of each fuzzing technique. For example, BFuzz, which showed the highest PR Ratio, mutated the dependent fields ($D$), resulting in test packets rejected by the target device.

One important observation is that a lower PR Ratio does not always indicate that a fuzzer is efficient. In the experimental results, we confirmed that the PR Ratio of Defensics was lower than that of \textsc{L2Fuzz}. This is mainly owing to following two reasons. First, Defensics exhibited a low rejection ratio because it hardly generated malformed packets. Furthermore, a Bluetooth application forms as many channels as the number of supported Bluetooth services. In \textsc{L2Fuzz}, some packets were rejected because \textsc{L2Fuzz} formed more channels than the maximum number in one L2CAP state. Because Defensics only tests one packet per state, there is less chance of being rejected. In summary, although Defensics showed a low PR Ratio, it hardly made malformed packets and did not sufficiently inspect each L2CAP state. 

In contrast, we confirmed that \textsc{L2Fuzz} showed a relatively low PR Ratio while generating a sufficiently large number of malicious packets, with the help of core field mutating technique.

\begin{table}[!h]
\caption{Results of the mutation efficiency measurement.}
\label{table:mt_eff}

\vspace{-0.5em}
\scalebox{1.15}{
\begin{tabular}{cccc}
\thickhline
\textbf{Fuzzer} & \textbf{MP Ratio} & \textbf{PR Ratio} & \textbf{Mutation efficiency} \\
\hline
L2Fuzz & 69.96\% & 32.49\% & 47.22\%  \\
Defensics & 2.38\% & 1.73\% & 2.33\% \\
BFuzz & 1.50\% & 91.60\% & 0.12\% \\
BSS & 0\% & 0\% & 0\% \\
\thickhline
\smallskip
\end{tabular}}
%\vspace{-6px}
\vspace{-1em}

\footnotesize{*MP Ratio = \textit{Malformed Packet Ratio}}\\
\footnotesize{*PR Ratio = \textit{Packet Rejection Ratio}}\\
\footnotesize{*Mutation efficiency = MP Ratio * (1 - PR Ratio)}\\

\vspace{-1.3em}
\end{table}

%\vspace{6px}
\textbf{Mutation efficiency measurement.}
We then calculated the mutation efficiency for each fuzzer using the measured MP and PR Ratios. \autoref{table:mt_eff} presents the measurement results.

%Although \textsc{L2Fuzz} showed a higher PR Ratio than Defensics, 
We confirmed that \textsc{L2Fuzz} was able to transmit the largest number of malformed packets without rejection; \textsc{L2Fuzz} showed a mutation efficiency of 47.22\%. The mutation efficiency value of Defensics, which showed the lowest PR Ratio, is 2.33\% because Defensics hardly generates malformed packets (\textit{i.e.}, MP Ratio was significantly low). Further, the mutation efficiency of BFuzz, which produced few malicious packets and showed a high rejection ratio, was 0.12\%, and the mutation efficiency of BSS, which failed to generate malicious packets, was 0\%. Moreover, \textsc{L2Fuzz} transmitted 524.27 packets per second (pps), allowing more packets to be tested in a shorter time than Defensics (3.37 pps), BFuzz (454.54 pps), and BSS (1.95 pps). 

From our experimental results, we confirmed that \textsc{L2Fuzz} outperformed existing Bluetooth fuzzing techniques in terms of generating more malformed packets that were less likely to be rejected by the target device.

\subsection{State coverage measurement} 
\label{state_cov}
Next, we examined the number of L2CAP states that each fuzzing technique could cover; the more covered states, the more likely the fuzzing technique to detect a Bluetooth vulnerability (see \autoref{phase2}).

\begin{figure} [h!]
\centering
    \includegraphics[width=\linewidth]{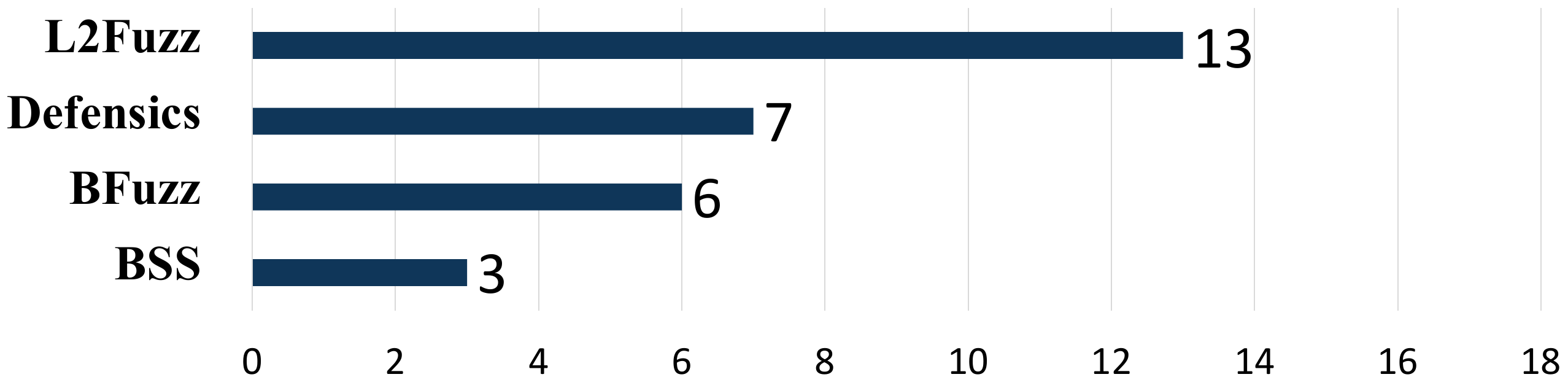}
\caption{L2CAP state coverage by different fuzzers.} 
\label{fig:state_coverage}
\vspace{1em}
\end{figure}

\begin{figure*} [t]
\centering
\includegraphics[width=\linewidth]{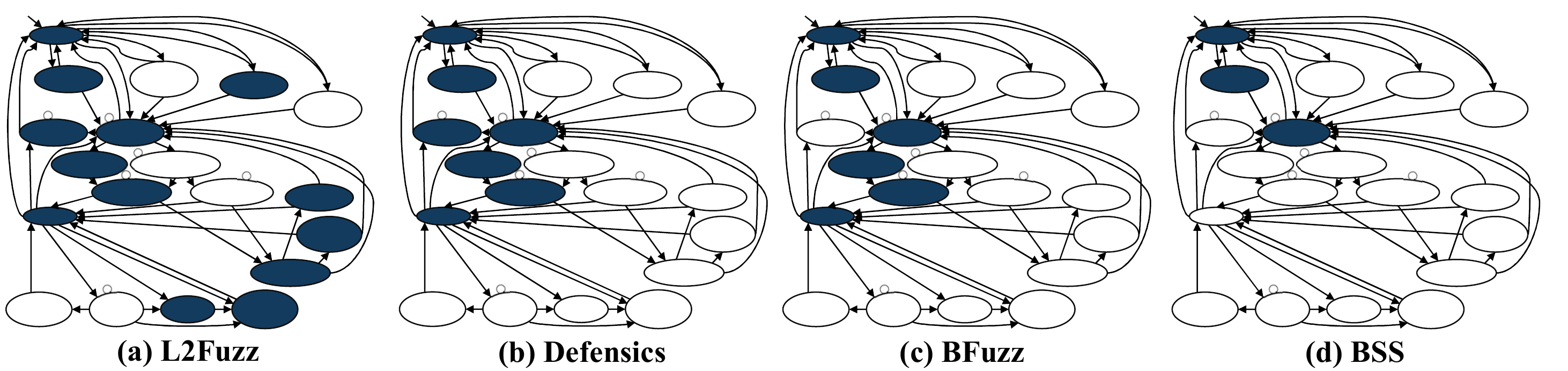}

\vspace{0.4em}
\caption{Illustration of the state coverage for each fuzzer based on the L2CAP state machine (see \autoref{fig:l2cap_state_machine}). 
Highlighted states represent testable L2CAP states in each fuzzer.}
%The highlighted state means a testable state.}
\label{fig:long}
%\vspace{1em}
\end{figure*}

We investigated each fuzzer's state coverage by analyzing the packet trace captured using PRETT~\cite{lee2018prett}. For fuzzers with fixed test times (\textit{i.e.}, Defensics), we analyzed the packet traces at the end of the test. For the remaining fuzzer with no test time limit, packet traces were analyzed at the end of a single test cycle. The results are shown in \autoref{fig:state_coverage} and \autoref{fig:long}.

%\vspace{6px}
\textbf{Results.}
From our experiment, we confirmed that \textsc{L2Fuzz} can cover almost twice as many L2CAP states (13 states) than existing fuzzers (at most seven states). \textsc{L2Fuzz} improved the accuracy of state transition by mapping only valid commands to each state, which rendered it possible to cover more L2CAP states (see \autoref{phase2}).
\textsc{L2Fuzz} could cover up to 13 states, including L2CAP states classified as \textit{move} and \textit{creation} jobs that were not covered by existing fuzzers. Conversely, the state coverage values of Defensics (\textit{i.e.}, seven states), BFuzz (\textit{i.e.}, six states), and BSS (\textit{i.e.}, three states) were less than that of \textsc{L2Fuzz} because they did not leverage valid commands for each state and were less effective at checking the target's response. One reason is that the Bluetooth specification document they used was outdated (\textit{i.e.}, Bluetooth core 2.1, published in 2007~\cite{old_bt_spec}). This is not at technical limitation, but it indicates that \textsc{L2Fuzz} is more efficient for checking Bluetooth devices that reflect the latest specifications.

In summary, \textsc{L2Fuzz} showed far superior mutation efficiency and state coverage compared to existing fuzzers. This indicates that \textsc{L2Fuzz} can detect L2CAP vulnerabilities of Bluetooth devices more effectively in practice.

\subsection{Case study}
\label{case_study}
We introduce a zero-day DoS vulnerability detected in Android Bluetooth devices (\textit{i.e.}, D2, see \autoref{table:devices}).

\begin{figure}[t]
\centering
    \includegraphics[width=1\linewidth]{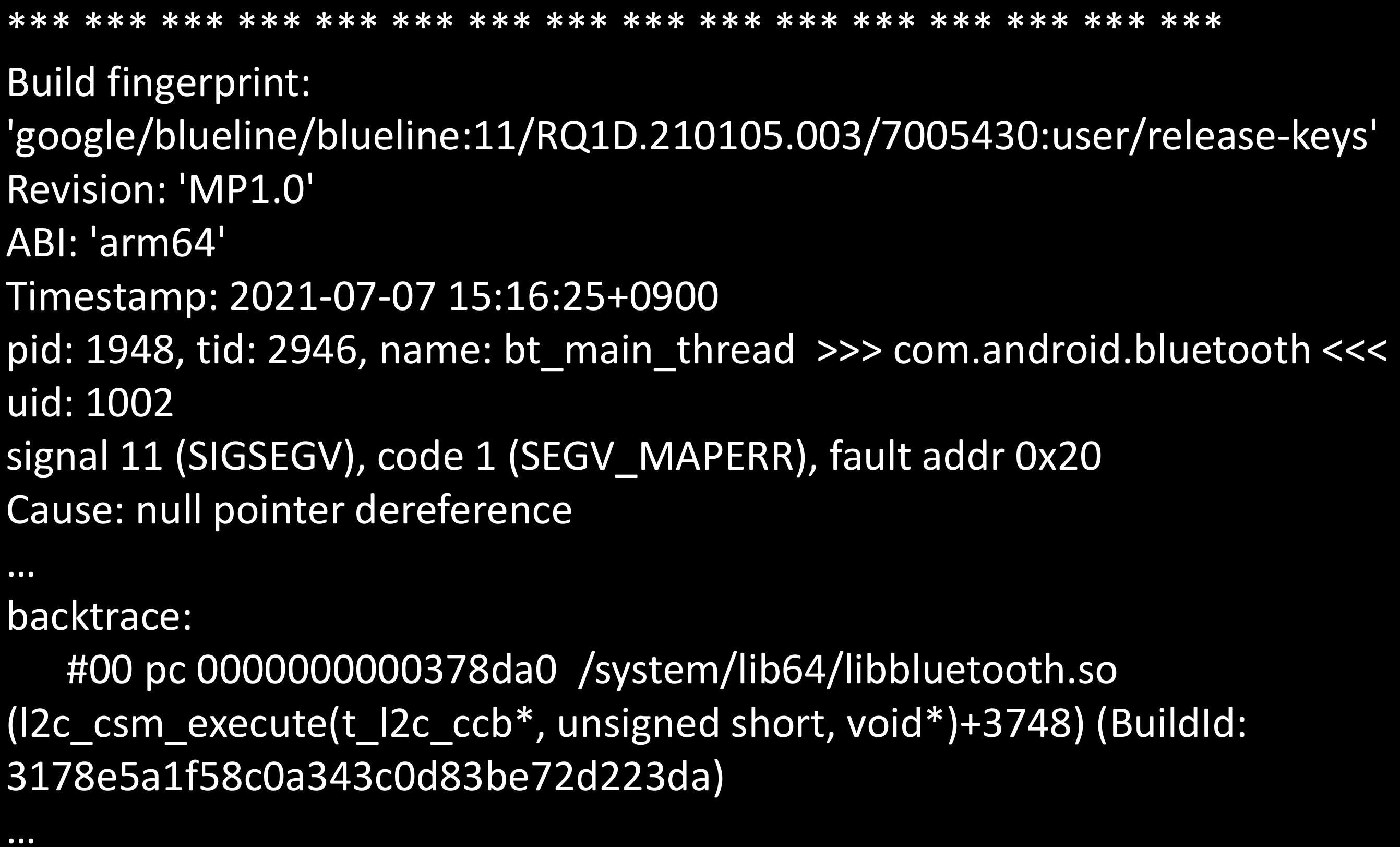}
\caption{Tombstone of Google Pixel 3 with BlueDroid. Vulnerability occurs at $t\_l2c\_ccb*$ (L2CAP channel control block) that uses  $\mathrm{CID_P}$.}
\label{fig:crash_log}
%\vspace{1em}
\end{figure}

\begin{figure}[t]
\centering
    \includegraphics[width=0.6 \linewidth]{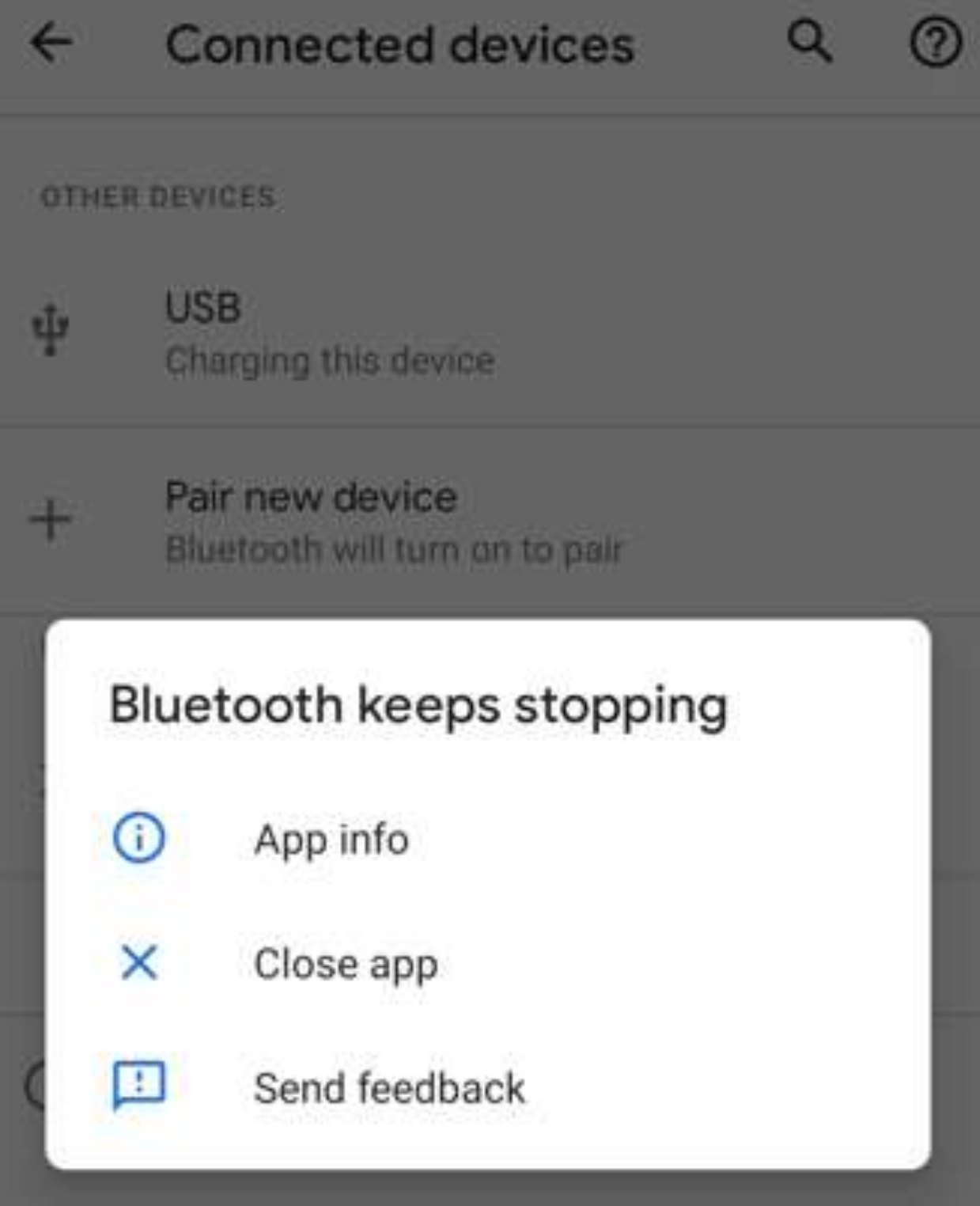}
\caption{Screenshot of crash message on Google Pixel 3. The device shows an error message and the Bluetooth is paralyzed.}
\label{fig:pixel3_crash}
\vspace{-1em}
\end{figure}

\textsc{L2Fuzz} connected to the %D3
D2 device's SDP port, and then performed state transition to the configuration states (\textit{i.e.}, configuration job). Afterwards, when a malicious packet with a DCID value of 0x40 and garbage added was sent to the target device, we confirmed that a null pointer deference was triggered in L2CAP layer of the target device. When a vulnerability occurs, the contents recorded in Tombstone~\cite{android_tombstone}, an Android crash dump file, are shown in \autoref{fig:crash_log}.

As a result of analyzing the root cause, the DCID and garbage values of the malformed packet influenced the channel control block of the L2CAP layer. In addition, the target device's display shows an error message about Bluetooth termination (see \autoref{fig:pixel3_crash}). To use Bluetooth again, we had to reset the Bluetooth function.
The \textsc{L2Fuzz} approach, which generates valid malformed packets while covering enough L2CAP states without pairing, could detect this zero-day vulnerability. Furthermore, DoS vulnerability was also detected on older versions of Android devices (\textit{i.e.}, D1 and D3). 
It is noteworthy that in the case of D3 (Galaxy7), DoS was triggered by malformed Create-Channel-Request that only \textsc{L2Fuzz} supports, and was detected in the Wait-Create state, which only \textsc{L2Fuzz} covers.

We responsibly reported this vulnerability to the Android security team. After discussing the cause and symptoms of the vulnerability, they became aware of the vulnerability (Android ID 195112457). % and agreed to apply the patch immediately, assigning a new CVE ID to the vulnerability.
Additionally,
we found a vulnerability that forced
Apple's wireless earphone device (\textit{i.e.}, D5, see \autoref{table:devices}) to shut down.
We reported this issue to the Apple security team, and they confirmed it and fixed the vulnerability \cite{apple_security_update}.

% shutting down Apple's wireless earphone device (D5, see \autoref{table:devices}) 
% and this was recognized by the manufacturer. A detailed description of this will be disclosed after obtaining the consent of the manufacturer.

%% file: 5-dis.tex
\section{Discussion}
\label{sec:dis}
Here we discuss several considerations related to \textsc{L2Fuzz} and countermeasures towards detected vulnerabilities.

\textbf{Applicability to other protocols.}
The methodology of \textsc{L2Fuzz} can be applied to other Bluetooth core protocols such as RFCOMM, SDP, and OBEX. 
Since these protocols also use their own state machines, we expect that the state guiding of \textsc{L2Fuzz} can lead users to test more states.
Also, the packet format of these protocols can be divided into core fields and other fields, thus we can apply the core field mutating technique.
Moreover, these protocols necessarily use L2CAP as they are on a higher layer than L2CAP (see \autoref{fig:bt_stack}).
This means that the generated \textsc{L2Fuzz}'s malformed packets (for testing L2CAP vulnerabilities) can also be used directly during fuzz testing for the protocols.
For these reasons, we determine that \textsc{L2Fuzz} can be applied to other Bluetooth protocols other than L2CAP; we leave this for future work.

\textbf{Countermeasures.}
To counter detected Bluetooth vulnerabilities, vendors of Bluetooth devices are encouraged to patch any detected vulnerabilities by updating the functionality that leverages PSM and $\mathrm{CID_P}$ in the Bluetooth L2CAP layer.
We also suggest encrypting each protocol as a fundamental solution.
Existing Bluetooth security technology has relied heavily on pairing. However, this cannot address attacks that do not require pairing, such as Blueborne (see \autoref{motivation}). Therefore, we believe that developing encryption methods for each protocol can resolve exposure to more vulnerabilities.

\textbf{Limitations and future work.}
First, although \textsc{L2Fuzz} effectively detected vulnerabilities in most cases, it was not capable of performing long-term fuzzing; when a fatal bug is triggered on the target device, it forcibly shuts down Bluetooth. Therefore, the tester must manually reset the device to perform another test. We will consider overcoming this issue by leveraging a virtual environment~\cite{ruge2020frankenstein}. Second,
\textsc{L2Fuzz} can detect vulnerabilities by analyzing the target's response packets; however, the root cause cannot be determined immediately.
We intend to resolve this issue by considering the internal log hooking that analyzes the crash root cause, similar to ToothPicker~\cite{heinze2020toothpicker}. Third, \textsc{L2Fuzz} cannot evaluate code coverage. Since Bluetooth devices are black-boxes and closed source, it is difficult to measure code coverage. We noted that Frankenstein~\cite{ruge2020frankenstein} succeeded in measuring code coverage in a limited way using binaries even though it required complex tasks such as firmware emulation. We will try to apply Frankenstein's method to \textsc{L2Fuzz.} Finally, while \textsc{L2Fuzz} covers a considerable number of L2CAP states, there are still cases where it does not, \textit{e.g.}, when \textsc{L2Fuzz} (as a master) connects with a target device that is a slave, there may be restrictions on the state that the target device can enter. We are considering leveraging techniques such as injecting applications that control state transitions of the test target.

%\textbf{Ethical considerations: Responsible vulnerability disclosure.}
\textbf{Responsible vulnerability disclosure.}
We reported all detected vulnerabilities in our experiments to the vendors: Android, Apple, Samsung, and the Ubuntu BlueZ team. Among them,
%a vendor-approved Android case that will be patched and assigned a CVE ID was introduced in \autoref{case_study}. Similarly, 
a crash found in Apple devices was patched by the vendors. The remaining vulnerabilities are currently under discussion. 
In addition, we have found several vulnerabilities in devices that are not mentioned in this paper. However, the information cannot be disclosed due to the vendor’s rejection.

%% file: 6-rel-works.tex
%\vspace{4px}
\section{Related work}
\label{sec:rel}
%In this section, we introduce several related studies.

%\vspace{6px}
\textbf{Bluetooth fuzzing techniques.} 
Existing Bluetooth fuzzing techniques (1) are inefficient for testing various Bluetooth devices, (2) do not generate valid malformed packets, and (3) do not cover enough L2CAP states.

Sweyntooth \cite{garbelini2020sweyntooth}, Frankenstein \cite{ruge2020frankenstein} and ToothPicker \cite{heinze2020toothpicker} attempted to detect Bluetooth vulnerabilities through fuzz testing. However, they did not focus on the Bluetooth BR/EDR host stack, which is a software commonly used in devices that provide Bluetooth services. In particular, Sweyntooth focused on Bluetooth Low Energy (BLE) protocol stack which is different from Bluetooth BR/EDR. Frankenstein focused on Bluetooth BR/EDR; however, it concentrates on the controller stack (firmware) that is different from the host stack (software). ToothPicker only focused on Apple's customized Bluetooth protocol stack, which is different from the common Bluetooth BR/EDR. 
Therefore, they are not suitable for detecting vulnerabilities in the commonly used BR/EDR host stack, which is the target of this paper.

There are several commercial Bluetooth fuzzers target BR/EDR host protocol stacks such as Bluetooth stack smasher (BSS) \cite{bss}, BFuzz \cite{kim2017poster} and Defensics \cite{defensics}. However, their test packets are not efficient in detecting vulnerabilities in Bluetooth devices (see \autoref{sec:eval}). Regarding BSS, it simply mutates only one field of a packet, which is insufficient to trigger vulnerabilities in the latest Bluetooth devices. BFuzz mutates packets that have previously been determined to be vulnerable; however, because it mutates almost every field, it is easily rejected by the target device. In the case of Defensics, most of the test packets are normal packets (i.e., not malformed packets); thus, instead of yielding unexpected behaviors, it often results in normal communication.

%\vspace{6px}
\textbf{Other Bluetooth vulnerability detection techniques.}
There are several approaches that attempted to detect Bluetooth vulnerabilities without using fuzz testing (\textit{e.g.}, KNOB~\cite{antonioli2019knob}, BIAS~\cite{antonioli2020bias} and BlueMirror~\cite{claverie2021bluemirror}). However, they are inefficient to test various devices because the scope of the target is limited and complicated implementations are required (\textit{e.g.}, they require link key sniffing, reverse engineering, and firmware patching). These tasks are difficult for the user to follow and implement; and also unsuitable for testing various Bluetooth devices.

% \vspace{6px}
%\textbf{General vulnerability detection techniques.}
%\vspace{6px}
\textbf{General vulnerability detection techniques.}
In addition, 
various approaches attempted to detect general vulnerabilities in a given codebase 
(\textit{e.g.}, \cite{kim2017vuddy, xiao2020mvp, woo2021centris}).
% there are several vulnerable code detection techniques that can be used to detect vulnerabilities in Bluetooth devices (\textit{e.g.}, VUDDY~\cite{kim2017vuddy} and MVP~\cite{xiao2020mvp}). 
Although these techniques can detect Bluetooth vulnerabilities, however, they can only be applied in an environment where the source codes of Bluetooth devices are available.
% Moreover, owing to the characteristics of the static analysis, they cannot detect vulnerabilities that occur only in running the devices. Therefore, their scopes are much limited than that of our proposed technique.

%% file: 7-concl.tex
\section{Conclusion}\label{sec:conclusion}
Security vulnerabilities in Bluetooth can pose a serious threat in the daily lives of people. In response, we present \textsc{L2Fuzz}, a stateful fuzzer for detecting
Bluetooth L2CAP vulnerabilities. By generating malformed packets (for testing purposes) that are less likely to be rejected by the target devices, \textsc{L2Fuzz} can detect potential vulnerabilities in Bluetooth devices more effectively than existing Bluetooth fuzzers. 
With \textsc{L2Fuzz}, developers can prevent risks in the Bluetooth host stack, which can increase the reliability of Bluetooth devices. 
The source code of \textsc{L2Fuzz} is available at \url{https://github.com/haramel/L2Fuzz} and 
will be publicly serviced at %the test tool will be updated at
\url{https://iotcube.net} as a part of BFuzz.

%that arise from the vulnerabilities

%% file: 8-appendix.tex
%\section{Android Denial-of-Service Tombestone.}
%\label{appendix_vul}

%% file: main.bbl
\begin{thebibliography}{10}
\providecommand{\url}[1]{#1}
\csname url@samestyle\endcsname
\providecommand{\newblock}{\relax}
\providecommand{\bibinfo}[2]{#2}
\providecommand{\BIBentrySTDinterwordspacing}{\spaceskip=0pt\relax}
\providecommand{\BIBentryALTinterwordstretchfactor}{4}
\providecommand{\BIBentryALTinterwordspacing}{\spaceskip=\fontdimen2\font plus
\BIBentryALTinterwordstretchfactor\fontdimen3\font minus
  \fontdimen4\font\relax}
\providecommand{\BIBforeignlanguage}[2]{{%
\expandafter\ifx\csname l@#1\endcsname\relax
\typeout{** WARNING: IEEEtran.bst: No hyphenation pattern has been}%
\typeout{** loaded for the language `#1'. Using the pattern for}%
\typeout{** the default language instead.}%
\else
\language=\csname l@#1\endcsname
\fi
#2}}
\providecommand{\BIBdecl}{\relax}
\BIBdecl

\bibitem{bt_market}
\BIBentryALTinterwordspacing
{Bluetooth SIG}. ({2021}) {Bluetooth market update}. [Online]. Available:
  \url{{https://www.bluetooth.com/ko-kr/bluetooth-resources/2021-bmu/}}
\BIBentrySTDinterwordspacing

\bibitem{defensics}
\BIBentryALTinterwordspacing
{Synopsys}. {Defensics Fuzz Testing}. [Online]. Available:
  \url{{https://www.synopsys.com/software-integrity/security-testing/fuzz-testing.html}}
\BIBentrySTDinterwordspacing

\bibitem{kim2017poster}
{Kim, Seulbae and Woo, Seunghoon and Lee, Heejo and Oh, Hakjoo}, ``{Poster:
  Iotcube: an automated analysis platform for finding security
  vulnerabilities},'' in \emph{{Proceedings of the 38th IEEE Symposium on
  Poster presented at Security and Privacy}}, {2017}.

\bibitem{bss}
\BIBentryALTinterwordspacing
{Pierre Betouin}. ({2006}, {May}) {[Infratech - release] version 0.6 de
  Bluetooth Stack Smasher}. [Online]. Available:
  \url{{http://www.secuobs.com/news/05022006-bluetooth10.shtml}}
\BIBentrySTDinterwordspacing

\bibitem{antonioli2019knob}
{Antonioli, Daniele and Tippenhauer, Nils Ole and Rasmussen, Kasper B}, ``{The
  KNOB is Broken: Exploiting Low Entropy in the Encryption Key Negotiation Of
  Bluetooth BR/EDR},'' in \emph{{28th USENIX Security Symposium (USENIX
  Security 19)}}, {2019}, pp. {1047--1061}.

\bibitem{antonioli2020bias}
{Antonioli, Daniele and Tippenhauer, Nils Ole and Rasmussen, Kasper}, ``{BIAS:
  bluetooth impersonation attacks},'' in \emph{{Proceedings of the 41st IEEE
  Symposium on Security and Privacy (SP)}}.\hskip 1em plus 0.5em minus
  0.4em\relax {IEEE}, {2020}, pp. {549--562}.

\bibitem{claverie2021bluemirror}
{Claverie, Tristan and Esteves, Jos{\'e} Lopes}, ``{BlueMirror: Reflections on
  Bluetooth Pairing and Provisioning Protocols},'' {2021}.

\bibitem{bt_spec}
\BIBentryALTinterwordspacing
{Bluetooth SIG}. ({2019}, {December}) {Bluetooth Core Specification 5.2}.
  [Online]. Available:
  \url{{https://www.bluetooth.com/ko-kr/specifications/bluetooth-core-specification/}}
\BIBentrySTDinterwordspacing

\bibitem{hua2008analysis}
{Hua, Yang and Zou, Yuexian}, ``{Analysis of the packet transferring in L2CAP
  layer of Bluetooth v2. x+ EDR},'' in \emph{{2008 International Conference on
  Information and Automation}}.\hskip 1em plus 0.5em minus 0.4em\relax {IEEE},
  {2008}, pp. {753--758}.

\bibitem{sharan2018air}
{Sharan, Ketan and Sharma, Neel and Sharda, Vangmayee and Arora, Neha}, ``{Air
  Mouse Using Bluetooth Technolgy},'' in \emph{{2018 Second International
  Conference on Intelligent Computing and Control Systems (ICICCS)}}.\hskip 1em
  plus 0.5em minus 0.4em\relax {IEEE}, {2018}, pp. {499--503}.

\bibitem{blueborne}
\BIBentryALTinterwordspacing
{Seri, Ben and Vishnepolsky, Gregory}. ({2017}, {November}) {BlueBorne: The
  dangers of Bluetooth implementations: Unveiling zero day vulnerabilities and
  security flaws in modern Bluetooth stacks.} [Online]. Available:
  \url{{https://www.armis.com/research/blueborne/}}
\BIBentrySTDinterwordspacing

\bibitem{garbelini2020sweyntooth}
{Garbelini, Matheus E and Wang, Chundong and Chattopadhyay, Sudipta and Sumei,
  Sun and Kurniawan, Ernest}, ``{Sweyntooth: Unleashing mayhem over bluetooth
  low energy},'' in \emph{{2020 USENIX Annual Technical Conference (USENIX ATC
  20)}}, {2020}, pp. {911--925}.

\bibitem{satam2017bluetooth}
{Satam, Shalaka Chittaranjan}, ``{Bluetooth Anomaly Based Intrusion Detection
  System},'' Ph.D. dissertation, {The University of Arizona}, {2017}.

\bibitem{prabadevi2014distributed}
{Prabadevi, B and Jeyanthi, N}, ``{Distributed Denial of service Attacks and
  its effects on Cloud Environment-a Survey},'' in \emph{{The 2014
  International Symposium on Networks, Computers and Communications}}.\hskip
  1em plus 0.5em minus 0.4em\relax {IEEE}, {2014}, pp. {1--5}.

\bibitem{dunning2010taming}
{Dunning, John}, ``{Taming the blue beast: A survey of bluetooth based
  threats},'' \emph{{IEEE Security \& Privacy}}, vol.~{8}, no.~{2}, pp.
  {20--27}, {2010}.

\bibitem{patel2012survey}
{Patel, Chandni M and Borisagar, APVH}, ``{Survey on taxonomy of ddos attacks
  with impact and mitigation techniques},'' \emph{{International Journal of
  Engineering Research and Technology}}, vol.~{1}, no.~{9}, {2012}.

\bibitem{yang2015cybersecurity}
{Yang, Yi and Jiang, HT and McLaughlin, Kieran and Gao, L and Yuan, YB and
  Huang, W and Sezer, Sakir}, ``{Cybersecurity test-bed for IEC 61850 based
  smart substations},'' in \emph{{2015 IEEE Power \& Energy Society General
  Meeting}}.\hskip 1em plus 0.5em minus 0.4em\relax {IEEE}, {2015}, pp. {1--5}.

\bibitem{torres2020nfdfuzz}
{Torres, George and Pesavento, Davide and Shi, Junxiao and Benmohamed, Lotfi},
  ``{NFDFuzz: A Stateful Structure-Aware Fuzzer for Named Data Networking},''
  in \emph{{Proceedings of the 7th ACM Conference on Information-Centric
  Networking}}, {2020}, pp. {169--171}.

\bibitem{afl}
\BIBentryALTinterwordspacing
{Michal Zalewski}. ({2017}, {April}) {American fuzzy lop}. [Online]. Available:
  \url{{https://github.com/google/AFL}}
\BIBentrySTDinterwordspacing

\bibitem{bluez}
\BIBentryALTinterwordspacing
{Bluetooth SIGBluetooth SIG"}. ({2021}) {Bluez : Official Linux Bluetooth
  protocol stack}. [Online]. Available: \url{{http://www.bluez.org/}}
\BIBentrySTDinterwordspacing

\bibitem{shafranovich2015bluetooth}
{Shafranovich, Yakov}, ``{Bluetooth data exchange between android phones
  without pairing},'' \emph{{arXiv preprint arXiv:1507.00650}}, {2015}.

\bibitem{jungbluemaster}
{Jung, Youngman and Shin, Junbum and Jang, Yeongjin}, ``{BlueMaster: Bypassing
  and Fixing Bluetooth-based Proximity Authentication}.''

\bibitem{nkuba2021riding}
{Carlos Kayembe Nkuba, Seulbae Kim and Sven Dietrich and Heejo Lee}, ``{Riding
  the IoT Wave With VFuzz: Discovering Security Flaws in Smart Homes},''
  \emph{{IEEE Access}}, vol.~{10}, pp. {1775--1789}, {2021}.

\bibitem{bluetooth_stack}
\BIBentryALTinterwordspacing
{Wikipedia}. ({2021}) {Bluetooth Stack}. [Online]. Available:
  \url{{https://en.wikipedia.org/wiki/Bluetooth\_stack}}
\BIBentrySTDinterwordspacing

\bibitem{muench2018you}
{Muench, Marius and Stijohann, Jan and Kargl, Frank and Francillon,
  Aur{\'e}lien and Balzarotti, Davide}, ``{What You Corrupt Is Not What You
  Crash: Challenges in Fuzzing Embedded Devices.}'' in \emph{{NDSS}}, {2018}.

\bibitem{lamping2004wireshark}
{Lamping, Ulf and Warnicke, Ed}, ``{Wireshark user's guide},''
  \emph{{Interface}}, vol.~{4}, no.~{6}, p.~{1}, {2004}.

\bibitem{lee2018prett}
{Lee, Choongin and Bae, Jeonghan and Lee, Heejo}, ``{PRETT: Protocol Reverse
  Engineering Using Binary Tokens and Network Traces},'' in \emph{{IFIP
  International Conference on ICT Systems Security and Privacy
  Protection}}.\hskip 1em plus 0.5em minus 0.4em\relax {Springer}, {2018}, pp.
  {141--155}.

\bibitem{android_tombstone}
\BIBentryALTinterwordspacing
{Google Android Security team}. ({2021}, {January}) {Debugging Native Android
  Platform Code}. [Online]. Available:
  \url{{https://source.android.com/devices/tech/debug}}
\BIBentrySTDinterwordspacing

\bibitem{old_bt_spec}
\BIBentryALTinterwordspacing
{Bluetooth SIG}. ({2007}, {July}) {Bluetooth Core Specification 2.1 + EDR}.
  [Online]. Available:
  \url{{https://www.bluetooth.com/ko-kr/specifications/specs/cs-core-specification-2-1edr/}}
\BIBentrySTDinterwordspacing

\bibitem{apple_security_update}
{Apple}. ({2021, December}) {Apple Security Update}.
  \url{https://support.apple.com/HT212975},
  \url{https://support.apple.com/HT212976},
  \url{https://support.apple.com/HT212978}, and
  \url{https://support.apple.com/HT212980}.

\bibitem{ruge2020frankenstein}
{Ruge, Jan and Classen, Jiska and Gringoli, Francesco and Hollick, Matthias},
  ``{Frankenstein: Advanced wireless fuzzing to exploit new bluetooth
  escalation targets},'' in \emph{{29th USENIX Security Symposium (USENIX
  Security 20)}}, {2020}, pp. {19--36}.

\bibitem{heinze2020toothpicker}
{Heinze, Dennis and Classen, Jiska and Hollick, Matthias}, ``{ToothPicker:
  Apple Picking in the iOS Bluetooth Stack},'' in \emph{{14th USENIX Workshop
  on Offensive Technologies (WOOT 20)}}, {2020}.

\bibitem{kim2017vuddy}
S.~Kim, S.~Woo, H.~Lee, and H.~Oh, ``Vuddy: A scalable approach for vulnerable
  code clone discovery,'' in \emph{Proceedings of the 38th IEEE Symposium on
  Security and Privacy (SP)}.\hskip 1em plus 0.5em minus 0.4em\relax IEEE,
  2017, pp. 595--614.

\bibitem{xiao2020mvp}
Y.~Xiao, B.~Chen, C.~Yu, Z.~Xu, Z.~Yuan, F.~Li, B.~Liu, Y.~Liu, W.~Huo, W.~Zou,
  and W.~Shi, ``Mvp: Detecting vulnerabilities using patch-enhanced
  vulnerability signatures,'' in \emph{29th USENIX Security Symposium (USENIX
  Security 20)}, 2020, pp. 1165--1182.

\bibitem{woo2021centris}
S.~Woo, S.~Park, S.~Kim, H.~Lee, and H.~Oh, ``{CENTRIS: A Precise and Scalable
  Approach for Identifying Modified Open-Source Software Reuse},'' in
  \emph{2021 IEEE/ACM 43rd International Conference on Software Engineering
  (ICSE)}.\hskip 1em plus 0.5em minus 0.4em\relax IEEE, 2021, pp. 860--872.

\end{thebibliography}
